\documentclass[]{aastex631}
\usepackage{CJK}
\usepackage{multirow}
\usepackage{fontenc}
\usepackage[figuresright]{rotating}
\usepackage{tabularx}
\usepackage{makecell}
\usepackage{amsmath}
\usepackage{booktabs}
\usepackage{url}
\received{----}
\revised{----}
\accepted{1 May, 2024}

\submitjournal{ApJ}

\begin{document}
\begin{CJK*}{UTF8}{gbsn}

\title{Modeling of Granulation in Red Supergiants in the Magellanic Clouds with the Gaussian Process Regressions}

\correspondingauthor{Yi Ren, Biwei Jiang}
\email{yiren@qlnu.edu.cn, bjiang@bnu.edu.cn}

\author[0000-0002-3828-9183]{Zehao Zhang (张泽浩)}
\affiliation{Institute for Frontiers in Astronomy and Astrophysics, Beijing Normal University, Beijing 102206, China}
\affiliation{Department of Astronomy, Beijing Normal University, Beijing 100875, China}

\author[0000-0003-1218-8699]{Yi Ren (任逸)}
\affiliation{College of Physics and Electronic Engineering, Qilu Normal University, Jinan 250200, China}

\author[0000-0003-3168-2617]{Biwei Jiang (姜碧沩)}
\affiliation{Institute for Frontiers in Astronomy and Astrophysics, Beijing Normal University, Beijing 102206, China}
\affiliation{Department of Astronomy, Beijing Normal University, Beijing 100875, China}

\author[0000-0002-7777-0842]{Igor Soszy\'nski}
\affiliation{Astronomical Observatory, University of Warsaw, Al. Ujazdowskie 4, 00-478 Warszawa, Poland}

\author[0000-0002-6244-477X]{Tharindu Jayasinghe}
\affiliation{Department of Astronomy, The Ohio State University, 140 West 18th Avenue, Columbus, OH, 43210, USA}




\begin{abstract}

The granulation of red supergiants (RSGs) in the Magellanic Clouds are systematically investigated by combining the latest RSGs samples and light curves from the Optical Gravitational Lensing Experiment and the All-Sky Automated Survey for Supernovae. The present RSGs samples are firstly examined for foreground stars and possible misidentified sources, and the light curves are sequentially checked to remove the outliers by white noise and photometric quality. The Gaussian Process regression is used to model the granulation, and the Markov Chain Monte Carlo is applied to derive the granulation amplitude $\sigma$ and the period of the undamped oscillator $\rho$, as well as the damping timescale $\tau$. The dimensionless quality factor $Q$ is then calculated through $Q=\pi \tau/\rho$. RSGs around $Q = 1/\sqrt{2}$ are considered to have significant granulation signals and are used for further analysis. Combining granulation parameters with stellar parameters, robust scaling relations for the timescale $\rho$ are established, while the scaling relations for amplitude $\sigma$ are represented by a piecewise function, possibly related to the tendency of amplitudes in faint RSGs to converge towards a certain value. Comparing results between the SMC and LMC confirms that amplitudes and timescales become larger with metallicity. In examining the scaling relations between the two galaxies, it is found that $\rho$ is nearly independent of metallicity, whereas $\sigma$ is more significantly affected by metallicity. The Gaussian Process method is compared with the periodogram fitting of the granulations, and the advantages of either are discussed.
\end{abstract}

\keywords{Red supergiant stars (1375); Stellar granulation (2102)}


\section{Introduction}  \label{sec:intro}

Red supergiants (RSGs) are noteworthy for their cool, luminous nature and status as evolved helium-burning stars. Most of the RSGs will explode as Type II-P supernovae, however, there is an absence of some brighter progenitors, which is known as the Red Supergiant Problem \citep{2009MNRAS.395.1409S, 2015PASA...32...16S, 2022MNRAS.515..897R}. There are currently several surveys targeting failed supernovae (e.g., \citet{2015MNRAS.450.3289G}), and at least one case, N6946-BH1, has long been considered a wonderful candidate \citep{2017MNRAS.468.4968A, 2024ApJ...962..145K}. The investigation of RSGs offers valuable insights into their evolution and their influential role in shaping the structure and evolutions of galaxies \citep{2024arXiv240206953R}. Among their many intriguing attributes, complex light variations, primarily driven by radial pulsation and convection, stand out. It is widely accepted that the light variations resulting from pulsations are rooted in the instability of radial fundamental, first-overtone, or even second-overtone modes \citep{1997A&A...327..224H, 2002ApJ...565..559G}, leading to period-luminosity relations akin to Cepheids \citep{2012ApJ...754...35Y, 2019ApJS..241...35R, 2024IAUS..376..292J}. The influence of convection is particularly pronounced in RSGs due to their substantial convective envelopes, where the motion of large convective cells frequently induces luminosity fluctuations \citep{1975ApJ...195..137S}.

The most direct manifestation of convection on a star's photosphere is the phenomenon known as granulation. Granulation was first observed by \citet{1801RSPT...91..265H} from the Sun, revealing a central area of hot and bright updraft surrounded by cooler and fainter downdraft regions at the edges. In particular, granulation of RSGs takes on a huge scale, with individual convective cells so large that they can cover a substantial fraction of the star's surface \citep{Norris_2021}. This is considered to be an explanation for the irregular light variations \citep{2020ApJ...898...24R}, which is supported by 3D hydrodynamic simulations \citep{2002AN....323..213F, 2010A&A...515A..12C}, interferometric \citep{2009A&A...508..923H} and spectropolarimetric observations \citep{2018A&A...620A.199L} of Betelgeuse. Another significant aspect of granulation in RSGs is its role in enhancing mass loss from these stars due to the movement of convective cells on the photosphere. This process leads to the generation of strong stellar winds, contributing substantially to the enrichment of dust in the vicinity of the star \citep{2018A&A...609A..67K, 2021AJ....161...98H}. Furthermore, it may even provide an explanation for the "great dimming" observed in Betelgeuse from 2019 to 2020, as proposed by \citet{2020ApJ...899...68D}.


Advancements in time-domain astronomy are developing a new era with a series of large-scale and long-duration surveys, including Convection, Rotation, and Planetary Transits (CoRoT; \citet{2006cosp...36.3749B}), Kepler \citep{2010Sci...327..977B}, the Optical Gravitational Lensing Experiment (OGLE; \citet{2015AcA....65....1U}), and the All-Sky Automated Survey for Supernovae (ASAS-SN; \citet{2017PASP..129j4502K}). These facilities yielded huge amount of stellar light curves, providing a valuable resource for the systematic exploration of stellar granulation. Granulation signals is thought to be "background noise" in asteroseismology, thus, efforts have been directed towards mitigating the influence of this noise \citep{1995A&A...293...87K}. However, an increasing number of studies begin to focus on this background signal and have investigated its relation with the physical properties of stars. The motivation for this change is due to the close relations of granulation parameters to stellar parameters. \citet{2011ApJ...741..119M} examined the light curves of approximately 1000 red giants (RGBs) observed by Kepler and derived their granulation parameters from the power spectrum. They conducted an analysis of the relationship between granulation parameters and stellar characteristics. In a larger sample of around 16,000 RGBs, \citet{2018ApJS..236...42Y} utilized long-cadence data from the Kepler mission to model solar-like oscillations and granulation signals. Nevertheless, it is worth noting that these previous studies predominantly concentrated on RGBs. In a recent study, \citet{2020ApJ...898...24R} introduced the observational study of granulation into RSGs for the first time. They employed the Continuous-time AutoRegressive Moving Average (CARMA) model to derive granulation parameters for RSGs in the Small Magellanic Cloud (SMC), Large Magellanic Cloud (LMC) and M31, and delved into their scaling relations with stellar parameters. Their work highlighted the potential of these scaling relations as a novel means of inferring stellar parameters from the light variations of stars. However, it is essential to note that the work by \citet{2020ApJ...898...24R} primarily encompassed the brighter RSGs, whose faint end is about $\log\ L/L_{\odot} = 4.3$ (see their Figures 9-10). With the expansion of the RSGs sample (i.e., about a triple increase) and the release of higher-quality light curves, there is an urgent need for further development in our understanding of the granulation characteristics of RSGs.

In practice, the granulation signal is inherently stochastic, intimately linked to convection, and typically manifests as a $1/f$ trend in the power spectral density (PSD), signifying a decrease in PSD with increasing frequency. To characterize this phenomenon, researchers often employ the Harvey function, originally introduced by \citet{1985ESASP.235..199H} while studying the Sun. \citet{1985ESASP.235..199H} noted that the autocovariance of granulation over time follows an exponential decay function defined by a characteristic timescale $\tau_{\mathrm{gran}}$ and variance $\sigma^2$, resulting in a Lorentz profile within the measured variation power spectrum:
\begin{equation}
     P(\nu)=\frac{4\sigma^2 \tau_{\mathrm{gran}}}{1+(2\pi \nu \tau_{\mathrm{gran}})^2}
     \label{equ: granulation}
\end{equation}
In fact, the exponent ``2" in the denominator characterizes the slope of the decay. In order to better fit the background power spectrum, one often chooses the exponent $\alpha$ as a free parameter \citep{2010MNRAS.402.2049H, 2014A&A...570A..41K, 2020ApJ...898...24R}, or a number of Harvey-like functions with different exponent values, e.g., 2 or 4 \citep{2009CoAst.160...74H, 2010A&A...522A...1K}. These fit functions can all be called modified Harvey functions, which are well-summarized by \citet{2011ApJ...741..119M}, and the trends in their results of these different methods are shown to be consistent (see Figures 2-4 of \citet{2011ApJ...741..119M}).

However, the observed PSD is influenced by windowing effects caused by the uneven sampling of the light curves, which can introduce bias in subsequent parameter estimates \citep{2018ApJS..236...16V, 2022A&A...668A.134B}. For ground-based observations, another factor is the periodicity of the sampling, since objects can only be observed at certain seasons of the year \citep{2009ApJ...698..895K}. Therefore, a more flexible and robust approach is needed to determine variability from the observed time series, which has been successfully used in the study of stochastic low-frequency (SLF) variability of upper main-sequence stars \citep{2022A&A...668A.134B} and the granulation and oscillation of RGBs \citep{2019MNRAS.489.5764P}, i.e., the Gaussian Process regression.

This study utilizes multi-band photometric time-series data obtained from OGLE and ASAS-SN to study the granulation of RSGs in the SMC and LMC using the Gaussian Process (GP) regression with \texttt{python} package {\sc celerite2} \citep{2017AJ....154..220F, 2018RNAAS...2...31F}. RSGs, being sufficiently luminous, are detectable in extragalactic galaxies. The Magellanic Clouds, as a nearby and extensively studied pair of dwarf irregular galaxies, serve as an excellent setting for the investigation of stellar populations and their characteristics. Moreover, the SMC and LMC, representing metal-poor galaxies in the Local Group, offer a valuable opportunity to explore RSG behavior within metal-poor environments. This paper aims to investigate the potential of employing the GP regressions to model granulation signals and to establish the relations between granulation and stellar parameters.

The paper is organized as follows: Section \ref{sec:data} provides an introduction to the RSG sample and the time-series data, including preprocessing steps. Section \ref{sec: GP} elucidates the mathematical aspects of GP regressions and the methodology for deriving granulation parameters. Section \ref{sec: scaling_relation} elaborates on the scaling relations between these granulation parameters and various stellar parameters. We discuss the metallicity effects and the methodology of PSD fitting in Section \ref{sec: discussion}. A concluding summary is presented in Section \ref{sec:summary}.

\section{Data and pre-processing} \label{sec:data}

\subsection{The RSGs sample} \label{sec:RSGs_sample}

Nnumerous collections of RSG samples have been carried out in the Magellanic Clouds because of their very close distances in comparison with other extragalaxies \citep{2002ApJS..141...81M, 2003AJ....126.2867M, 2006ApJ...645.1102L, 2015A&A...578A...3G, 2019A&A...629A..91Y, 2021A&A...646A.141Y, 2023ApJ...959..102D}. In all of these literatures, one of the biggest challenges is to eliminate the contamination of foreground red dwarf stars. An effective way is to make use of the excellent astrometric measurements by Gaia (e.g., \citet{2019A&A...629A..91Y, 2021A&A...646A.141Y}), Recently, \citet{2021ApJ...923..232R} removed more foreground stars according to the obviously distinct positions of dwarf and giant stars in the near-infrared color-color diagram $(J-H)_0/(H-K)_0$. With the updated {\em Gaia}/EDR3, they identified a total of 2,138 and 4,823 RSGs in the SMC and LMC, respectively, in the near-infrared color-magnitude diagram (CMD). This constitutes the most extensive sample of RSGs in the Magellanic Clouds thus far and serves as the initial sample for this work.

\begin{figure}[h]
	\centering
    \includegraphics[scale=0.36]{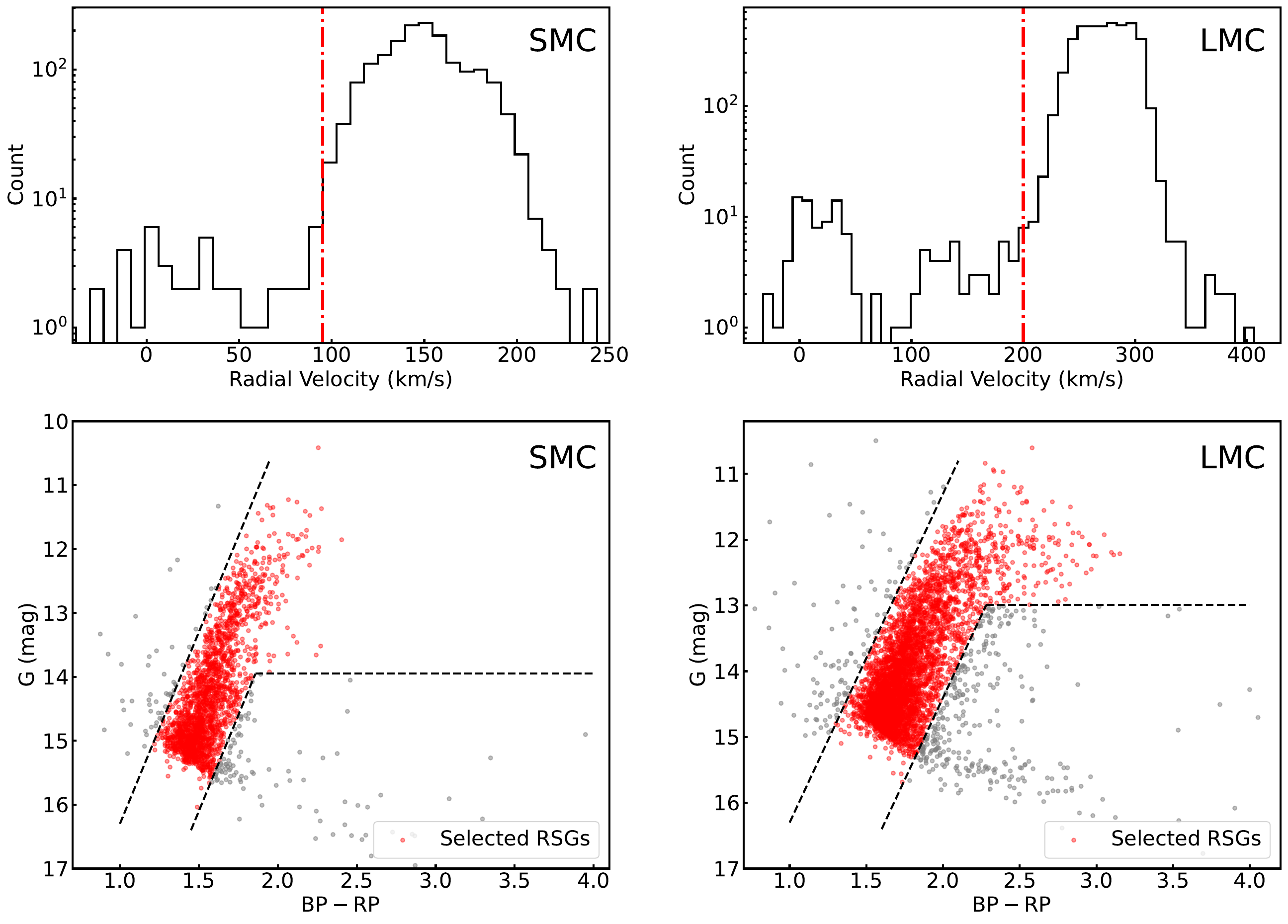}
	\caption{Removal of the foreground stars by the radial velocity (the upper panels) and the non-RSG stars by the {\em Gaia} color-magnitude diagram $BP-RP$ vs. $G$ (the lower panels) for the SMC (the left panels) and LMC (the right panels) respectively. The red lines in the upper panels mark the value of 95 km/s and 200 km/s for SMC and LMC respectively, and the red dots in the bottom panels denote the selected RSGs, while the gray dots are removed sources, and the black dashed lines are the boundaries of the RSGs. \label{fig:Gaia_RV_CMD}}
\end{figure}

Considering that \citet{2021ApJ...923..232R} used parallax and proper motion to eliminate the foreground dwarfs of the SMC and LMC, we conduct an additional removal of foreground sources using the radial velocity (RV) parameter. The RV data from {\em Gaia}/DR3 \citep{2023A&A...674A...5K} is selected first for its coverage of about 90\% sample stars. Some of the stars without {\em Gaia} data on RV, the APOGEE measurement is searched and used since there is a good agreement of RV between APOGEE and {\em Gaia} \citep{2022ApJS..259...35A}. Only sources with RV \textgreater\ 95 km/s in the SMC and RV \textgreater\ 200 km/s in the LMC are retained. Specifically, there are 4,592 stars with the {\em Gaia} RV and additional 39 stars with the APOGEE RV values among all the 4,823 RSG candidates in LMC, among which 123 and 10 stars respectively are removed by the criterion. For the SMC, 1691 stars has the {\em Gaia} RV values and 44 of them are removed, meanwhile additional 6 stars have the APOGEE RV value and none is removed. To ensure our sample is reliably identified in the optical bands, the {\em Gaia} CMD is taken to examine the reliability of the RSGs. The RSGs boundary on the {\em Gaia} CMD is determined by two diagonal lines and a horizontal line, as shown in Figure \ref{fig:Gaia_RV_CMD}. The horizontal lines are selected from \citet{2019A&A...629A..91Y} and \citet{2021A&A...646A.141Y}, remarking $G=13.95$ for the SMC and $G=12.99$ for the LMC, respectively, which indicate the  upper limit of luminosity of asymptotic giant branch stars (AGBs). This selection lead to the further removal of 153 and 424 stars from the SMC and LMC, respectively. The stars excluded are mainly foreground dwarfs, which have a contamination rate of about 2\%, and AGBs which can clearly be seen on the CMD in Figure \ref{fig:Gaia_RV_CMD} - fainter than RSGs with a tendency to extend towards the red end. Besides, the bluer stars on {\em Gaia} CMD are most likely RSG binaries with a hotter companion, which are not sensitive in the near-infrared band \citep{2018AJ....156..225N, 2019ApJ...875..124N, 2020ApJ...900..118N}. After all these exclusions, 1,941 and 4,266 RSGs are finally selected from the initial samples in SMC and LMC, respectively. Table \ref{tab:RSG_numbers} present the rules and results of each selection process in details.

\subsection{The time-series data}

\subsubsection{OGLE} \label{sec:ogle}

The OGLE observation uses the 1.3 m Warsaw telescope, located at the Las Campanas Observatory in Chile \citep{2015AcA....65....1U}. The telescope is equipped with a 32-CCD detector Mosaic camera and covers about 1.4 square degrees of the sky. This study based on the fourth phase of the OGLE survey (OGLE-IV) $I$-band light curves collected between 2010 and 2020 from the SMC and LMC. These observations, with a cadence ranging from 4 to 8 days, span an approximate duration of 3600 days. Over this period, a typical light curve features between 600 to 900 observational points. Furthermore, the precision of these observations is notably high, with the photometric uncertainty of nearly all observations being 0.005 magnitudes, allowing for highly accurate characterizations of stellar variability.

\subsubsection{ASAS-SN} \label{sec:asas-sn}

ASAS-SN is a ground-based survey that covers the entire sky. It started with four telescopes in Hawaii and Chile, monitoring the sky on the cadence of 2-3 days at a depth of about $V=17$ magnitude. Until now, the ASAS-SN project has expanded to 20 telescopes around the world, monitoring the g-band sky at a depth of 18.5 magnitude on a cadence of ~1 day. For the $g$-band observations, the number of photometric measurements per light curve typically clusters around 700, 1200, and 1600, while for the $V$-band, the numbers are more concentrated around 500 and 800 measurements. The difference in photometric point distribution are attributed to the spatial location of the stars being observed. The observational duration for the $g$-band extends to about 1880 days, whereas for the $V$-band, it is approximately 1600 days. We collected both $V$-band and $g$-band light curves, applying corrections for the zero-point offsets between the different cameras \citep{2018MNRAS.477.3145J} and recalibrating the photometric errors \citep{2019MNRAS.485..961J}. Regarding the photometric uncertainties, for stars with $g<16$ magnitude, the average photometric uncertainty is less than 0.08 magnitudes. Similarly, for stars brighter than $V=15$ magnitude, the uncertainty falls below 0.05 magnitudes.

\subsection{Data pre-processing}

\subsubsection{Light curve outliers}
\begin{figure}[h]
	\centering
    \includegraphics[width=\textwidth]{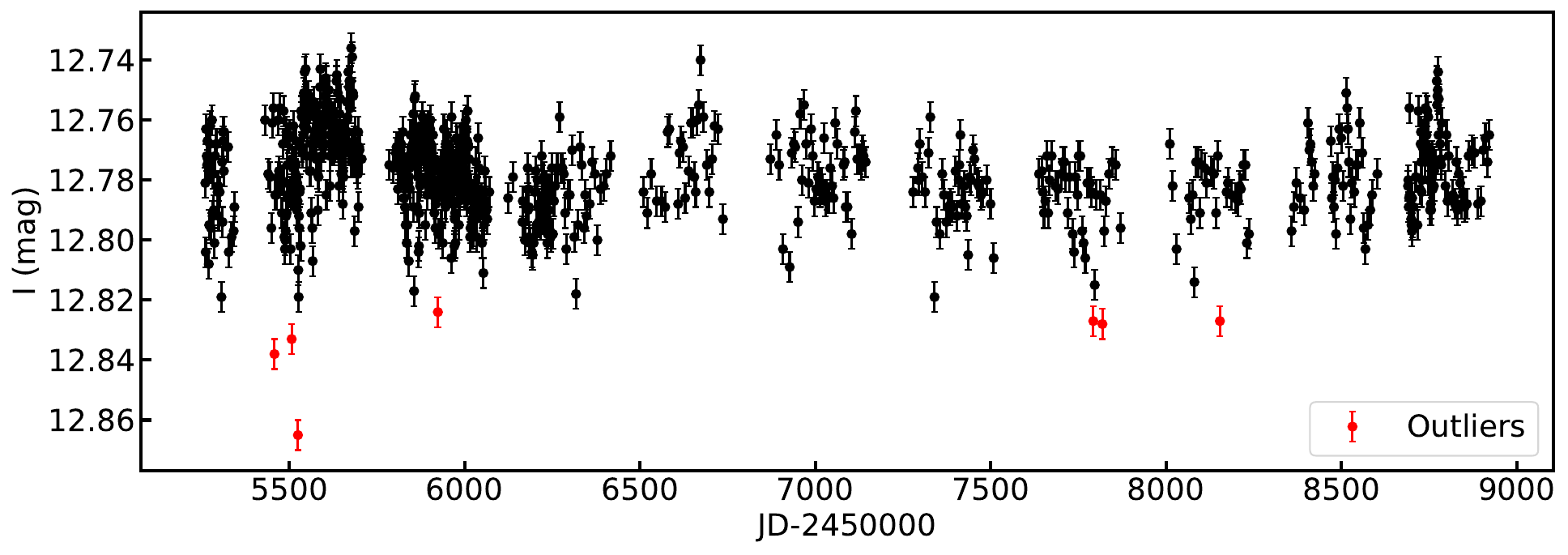}
	\caption{An example of the OGLE $I$-band light curve (ID: LMC518.09.13530) processing. The black dots represent the retained photometric measurements, while the red dots, identified as outliers, are removed. \label{fig: LC_Outliers}}
\end{figure}

The collected light curves are processed further. We carry out a simple but effective way to remove outliers in light curves. For each photometric point $X_{i}$, if $X_{i} - \bar{X} < 3\sigma_{X_{i}}$, it is considered to be an outlier and is removed, where $\bar{X}$ and $\sigma_{X_{i}}$ represent the mean magnitude and the standard deviation of a light curve, respectively. Figure \ref{fig: LC_Outliers} shows an example where points that clearly deviate from the normal range are removed.

\subsubsection{White noise}\label{sec: white_noise}
The white noise test is further applied to these data. The motivation is that the ASAS-SN light curves come from forced photometry of a given RSGs coordinate, though the brightness of some faint sources is fainter than the detection threshold, a light curve is still generated. It is well known that white noise is a signal independent of frequency with a constant PSD. To test whether a time-series photometric results can be well modeled by a white noise, its sample autocorrelation function (ACF) is calculated. Specifically, if more than 90\% of ACFs do not exceed the range of $\pm 1.96 / \sqrt{L}$, the sequence is considered to be a white noise and eliminated, where $L$ is the length of the sequence, and 1.96 represents the z-value of the 95\% confidence interval in a normal distribution \citep{Fisher1992}. If 84\%-90\% of ACFs are within this range, the results of Kolmogorov-Smirnov test are further taken into account, that is, if the sequence follows the normal distribution, it is also considered as white noise. In total, there are 1041, 1476, and 33 light curves considered to be white noise in the $g$, $V$, and $I$ bands of the SMC, and 1736, 2357, and 73 light curves in the $g$, $V$, and $I$ bands of the LMC, respectively. Light curves that are considered white noise are eliminated and the remaining light curves are used for further analysis.

\subsubsection{Photometric quality}
\begin{figure}[h]
	\centering
    \includegraphics[width=\textwidth]{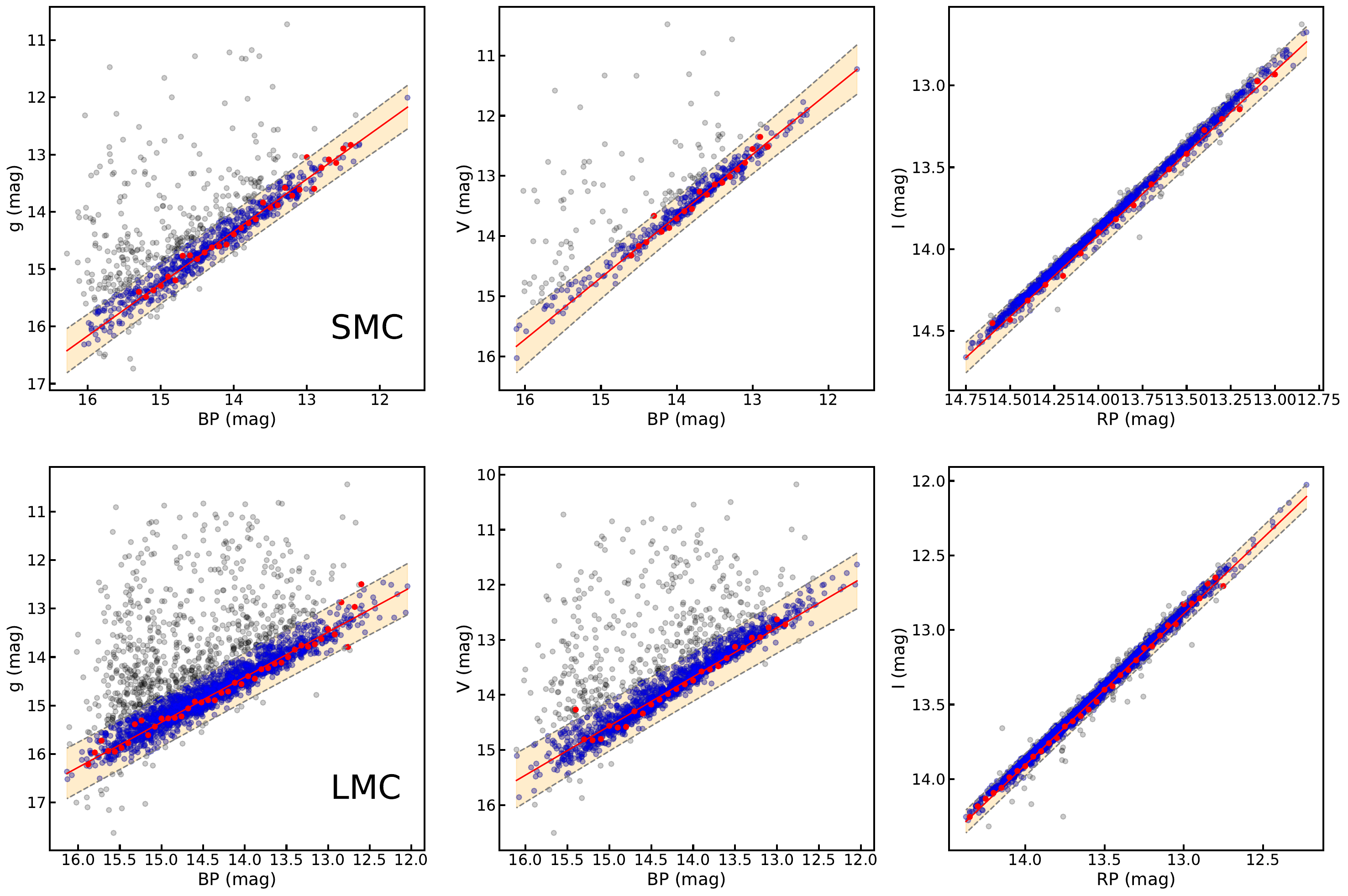}
	\caption{Comparison of the mean magnitude of the $g$, $V$ and $I$-band light curves with the {\em Gaia} $BP$ or $RP$ magnitude for the SMC (the upper panels) and the LMC (the lower panels). The red dots denote the position with the densest distribution in the Y-axis in each bin, and the red lines denote the fitting lines of red dots with the light orange shadow areas being 99\% confidence interval of the linear fittings (shown in blue dots). \label{fig: Photomtric_selection}}
\end{figure}

To evaluate the photometric reliability of the remaining light curves, the average magnitudes are compared with the {\em Gaia} magnitudes. The ASAS-SN $g$ and $V$ magnitudes are compared with the $BP$ magnitude and the OGLE $I$ magnitude with the $RP$ magnitude. Given the similar effective wavelengths of these bands, a close agreement is anticipated between these magnitudes. Figure \ref{fig: Photomtric_selection} shows the comparison, where most of the sources exhibit very good agreement between the ground-based observation and the {\em Gaia} space observation, while some sources deviate apparently from the agreement line, in particular in the ASAS-SN $g$ and $V$ bands. The outliers are examined in the Simbad photometric images and they are found to locate in crowded star fields (e.g., star clusters) and very difficult to separate from the neighbouring stars by the ground-based observation. On the other hand, the {\em Gaia} space observation has much higher spatial resolution and is able to isolate the object precisely. Consequently, the photometry can easily lead to an over-estimation of stellar brightness in the visual bands for the ground observation, which is right the phenomenon present in Figure \ref{fig: Photomtric_selection}. In contrast, the $I$-band photometric quality is much better mainly because RSGs are dominantly brighter than the other stars in the infrared band. The stars with poor photometric quality are removed through calculating the mean relation and its scattering.  The {\em Gaia} magnitude (X-axis) is sliced to find the position with the densest distribution of the mean magnitude of the light curves (Y-axis) in each bin, and then a linear fitting is performed to the ridge, as shown by the red dots and red lines in Figure \ref{fig: Photomtric_selection}. The 99\% confidence interval of the linear fittings is used to delimit the range of accepted sources. Through such selection, 298, 98 and 29 light curves are removed from $g$, $V$ and $I$ bands of the SMC, and 684, 385 and 63 light curves are removed from $g$, $V$ and $I$ bands of the LMC, respectively.

The results of the above elimination are presented in Table \ref{tab: LC_number}.

\section{Inferring granulation parameters with Gaussian Process regression} \label{sec: GP}

\subsection{Gaussian Process regression}\label{sec: GP_regression}

Currently, several common methods for studying stochastic variations include the fitting of PSD \citep{2011ApJ...741..119M, 2014A&A...570A..41K, 2020ApJ...896..164D, 2020ApJ...902...24D}, GP regression \citep{2019MNRAS.489.5764P, 2022A&A...668A.134B}, and CARMA models \citep{2020ApJ...898...24R}. GPs, as a method for directly fitting time series data, avoid the transformation of light curves into Fourier space, thereby circumventing potential biases in fitting PSD mentioned in Section \ref{sec:intro}. As for the CARMA model, a particular challenge is the selection of the model order (see Section 3.2 of \citet{2020ApJ...898...24R}), as it is not possible to know at which order the observed stochastic variations occur, thus the inferred parameters are greatly constrained by the model, or alternatively, a mixture model might be used \citep{2018ApJ...869..178T}. GPs describe each data point in the time series as correlated random variables with a mean value and a variance, where any finite set of these variables follows a multivariate Gaussian distribution \citep{2019MNRAS.489.5764P, 2023ARA&A..61..329A}. Therefore, correlated stochastic variations in the time series can be described by defining the covariance matrix.

The most crucial aspect of GPs is the selection of a kernel that describes the correlations among time series. This kernel is then used to find the set of parameters that best reproduces the observed data. {\sc celerite2} \citep{2017AJ....154..220F, 2018RNAAS...2...31F} offers a kernel in the form of a sum of exponential functions: 
\begin{equation}
     k_{\alpha}(\tau_{ij}) = \sum_{m=1}^{M} a_m \exp(-c_m \tau_{ij}),
\end{equation}
where $\alpha=\{a_m,c_m\}$, $\tau_{ij} = |t_i - t_j|$ represent the absolute distance in time series between points $i$ and $j$. Specifically, to describe stellar variability, a kernel based on a stochastically driven, damped harmonic oscillator (SHO) is provided, whose PSD takes the following form:
\begin{equation}
     S(\omega) = \sqrt{\frac{2}{\pi}} \frac{S_0 \omega^4_0}{\left(\omega^2 - \omega^2_0\right)^2 + \omega^2\omega^2_0 / Q^2},
     \label{equ: SHO}
\end{equation}
where $\omega$ is the angular frequency, $\omega_0$ is the frequency of the undamped oscillator, $Q$ is the quality factor, and $S_0$ is the parameter related to the variability amplitude. As stated in \citet{2017AJ....154..220F}, the kernel whose PSD is Equation \ref{equ: SHO} is given by 
\begin{equation}
k(\tau; S_0, Q, \omega_0) = S_0 \omega_0 Q e^{-\frac{\omega_0 \tau}{2Q}} \times
\begin{cases}
\cosh(\eta \omega_0 \tau) + \frac{1}{2\eta Q} \sinh(\eta \omega_0 \tau), & 0 < Q < \frac{1}{2}, \\
2(1 + \omega_0 \tau), & Q = \frac{1}{2}, \\
\cos(\eta \omega_0 \tau) + \frac{1}{2\eta Q} \sin(\eta \omega_0 \tau), &  Q > \frac{1}{2} ,
\end{cases}
\end{equation}
where $\eta = \vert 1- (4Q^2)^{-1}\vert^{1/2}$. \citet{2017AJ....154..220F} discussed several limits of physical interest, it is worth mentioning that for $Q=1/\sqrt{2}$, Equation \ref{equ: SHO} simplifies to 
\begin{equation}
	S(\omega) = \sqrt{\frac{2}{\pi}} \frac{S_0}{(\omega/\omega_0)^4+1},
\end{equation} 
which has the similar form of Equation \ref{equ: granulation}, and is commonly used to model the granulation signal (e.g., \citep{2019MNRAS.489.5764P} and \citep{2020ApJ...898...24R}).

\subsection{Estimation of granulation parameters}

Inspired by \citet{2019MNRAS.489.5764P} and \citet{2022A&A...668A.134B}, we select a SHO kernel to model the granulaiton signal in RSGs. \citet{2022A&A...668A.134B} discussed two approaches to modeling: one treats $Q$ as a free parameter, while the other fixes $Q = 1/\sqrt{2}$. They concluded that allowing $Q$ to vary freely offers a more robust method, as it introduces additional complexity into the model, enabling a more flexible and accurate representation of the time series variability. Following them, we build up our SHO kernel with three parameters: $\sigma$, $\rho$, and $\tau$, with $\sigma = \sqrt{S_0 \omega_0 Q}$, $\rho = 2 \pi / \omega_0$, $\tau = 2Q/\omega_0$, representing alternative parameters of Equation \ref{equ: SHO}. Moreover, to address the white noise in the Fourier space of the time series, we add a constant, $C_W$, in quadrature to the diagonal of the covariance matrix as jitter. Its kernel function is: 
\begin{equation}
	k(t_i, t_j) = C_{\rm jitter}^{2} \delta_{i,j},
\end{equation}
where $C_{\rm jitter}^{2}$ is the GP regression jitter term.

A direct fitting of these parameters is performed with {\sc lmfit} \citep{matt_newville_2018_1699739} to determine the initial values of these parameters, followed by Monte Carlo Markov Chain (MCMC) sampling using {\sc emcee} \citep{2013PASP..125..306F} to explore their posterior distributions. We conduct sampling in the logarithmic space and set non-informative Gaussian priors for each parameter, centered at zero with a standard deviation of 2.0 \footnote{As suggested in \url{https://celerite2.readthedocs.io/}.}. This choice reflects a balance between allowing flexibility in the values of parameters and penalizing extreme deviations, thus fostering the exploration of parameter space around physically plausible values. The sampler underwent an initial burn-in phase followed by a subsequent run with 10000 steps to achieve the accurate parameter estimation, i.e., the 50th percentile. The uncertainties come from the 16th and 84th percentiles of the posterior probability distribution of the parameters.

\begin{figure}[htb]
	\centering
     \includegraphics[width=0.6\textwidth]{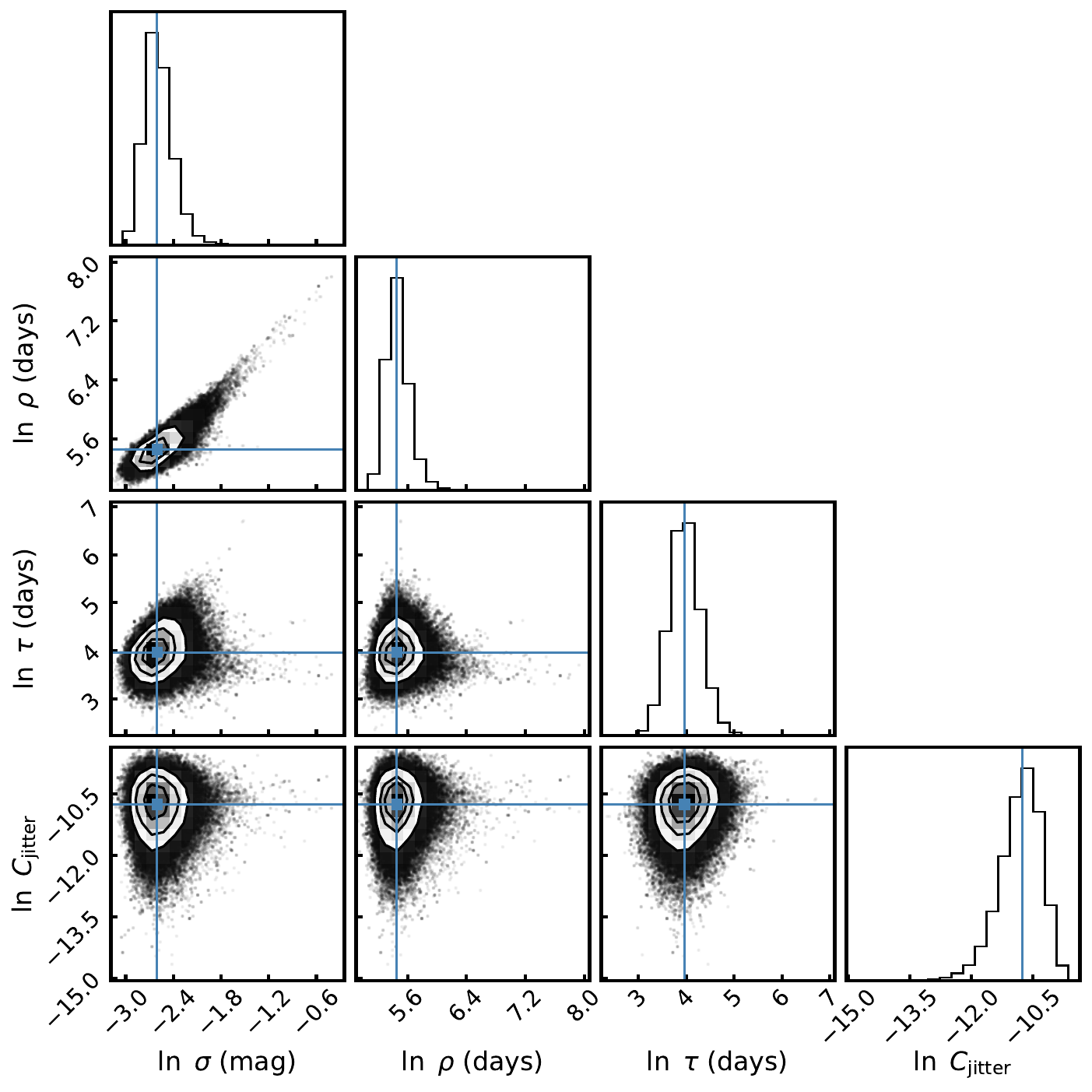}
	\caption{The corner plot presenting the joint parameter distributions of $\ln \sigma$, $\ln \rho$, $\ln \tau$ and $\ln C_{\rm jitter}$ of ASAS-SN $V$-band light curve (ID: 10024). Each panel on the diagonal represents the individual probability density functions for the corresponding parameter, while the off-diagonal panels illustrate the two-dimensional projections, revealing the correlations and degeneracies among the parameters. \label{fig: corner}}
\end{figure}

Goodness of fittings is ensured by the convergence of the chains. The Gelman-Rubin (GR) diagnostic \citep{1992StaSc...7..457G} is used to determine if the fittings are converged. The GR statistic (denoted as R-hat, $\hat{R}$) quantifies the convergence of MCMC simulations by comparing the variance between different chains to the variance within each chain, thus reflecting the consistency of parameter estimations across multiple runs. Although a widely used threshold is $\hat{R} < 1.1$, it is controversially considered too high to yield reasonable estimates of target quantities (see \citet{2018arXiv181209384V} for example). Thus, a cufoff of $\hat{R} < 1.01$ \citep{2019arXiv190308008V, 2023ApJ...945..132C} is adopted. The majority of the fittings meets the aforementioned criterion; specifically, within the SMC, 553 (92\%) in the $g$-band, 351 (96\%) in the $V$-band, and 1177 (97\%) in the $I$-band light curves converged; whereas in the LMC, 1744 (94\%) in the $g$-band, 1461 (96\%) in the $V$-band, and 2262 (98\%) in the $I$-band shows convergence. Figure \ref{fig: corner} shows an example of posterior probability distributions of $\ln \sigma$, $\ln \rho$, $\ln \tau$ and $\ln C_{\rm jitter}$.

Consequently, we obtain all the necessary parameters. The parameter $\sigma$ represents the root mean square (RMS) amplitude of the brightness fluctuations driven by granulation, measured in magnitudes, while $\rho$ denotes the characteristic timescale of granulation, measured in days. These definitions are consistent with Equation 12 from \citet{2019MNRAS.489.5764P}. Following the earlier definition in this section, the dimensionless quality factor $Q=\pi \tau/ \rho$ can be calculated, which characterizes the general properties of the variability. When $Q$ is large, the kernel function is used to describe undamped, high-quality oscillation, indicating less stochasticity and stronger quasi-periodicity in the light curves. A more detailed discussion about $Q$ will be presented in the following subsection.

\subsection{The bimodal distribution of $Q$}\label{sec: Q_distribution}

\begin{figure}[h]
	\centering
    \includegraphics[width=\textwidth]{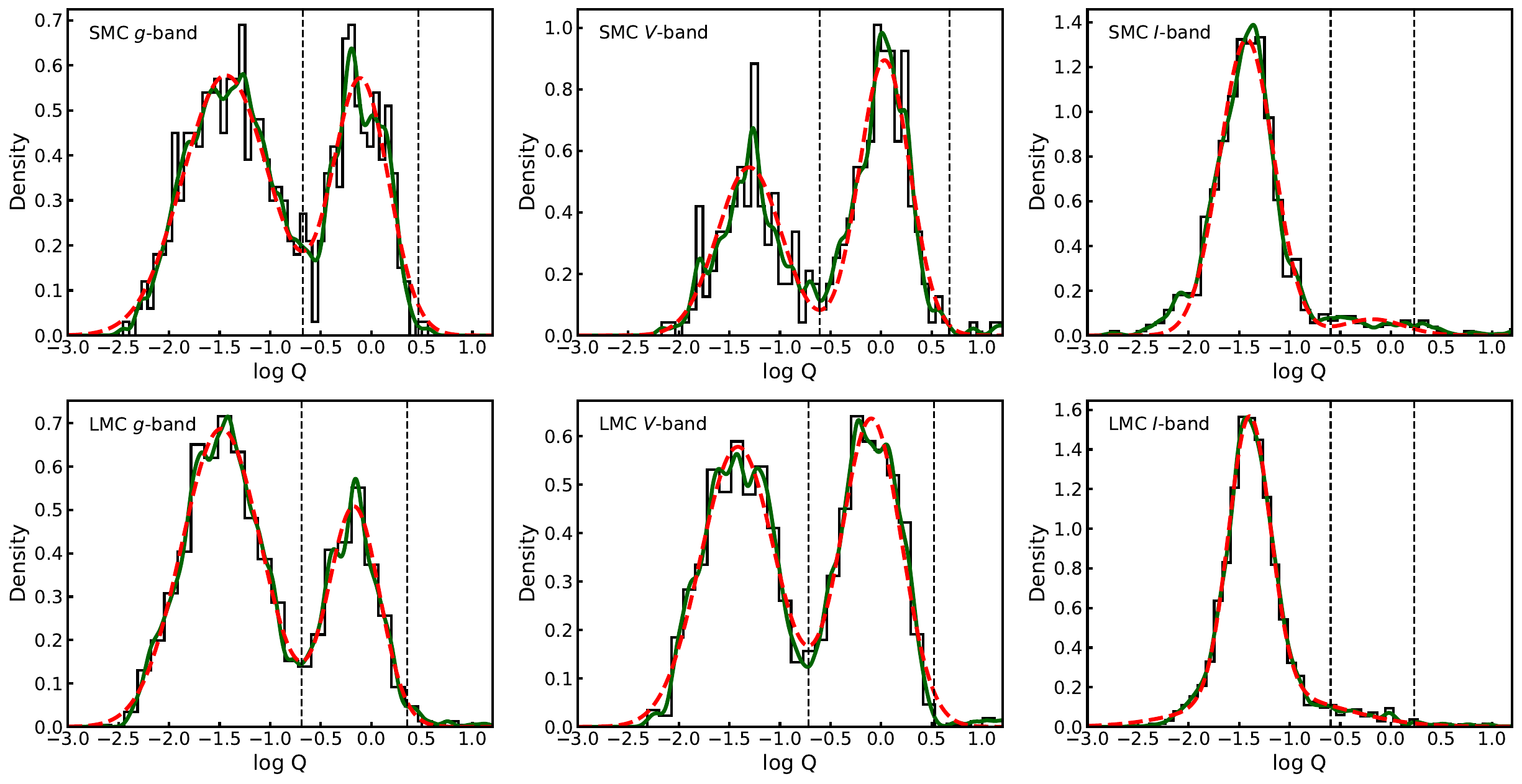}
	\caption{The distribution of $Q$ and its double-Gaussian fitting. The upper and lower panels display the results for the SMC and LMC, respectively, with the $g$, $V$, and $I$ bands presented from left to right. The green solid line represents the kernel density estimation, and the red dashed line illustrates the double-Gaussian fitting. The black dashed lines indicate the selected range of $Q$ used for further analysis.} \label{fig: Q_distribution}
\end{figure}

\begin{figure}[h]
	\centering
    \includegraphics[width=\textwidth]{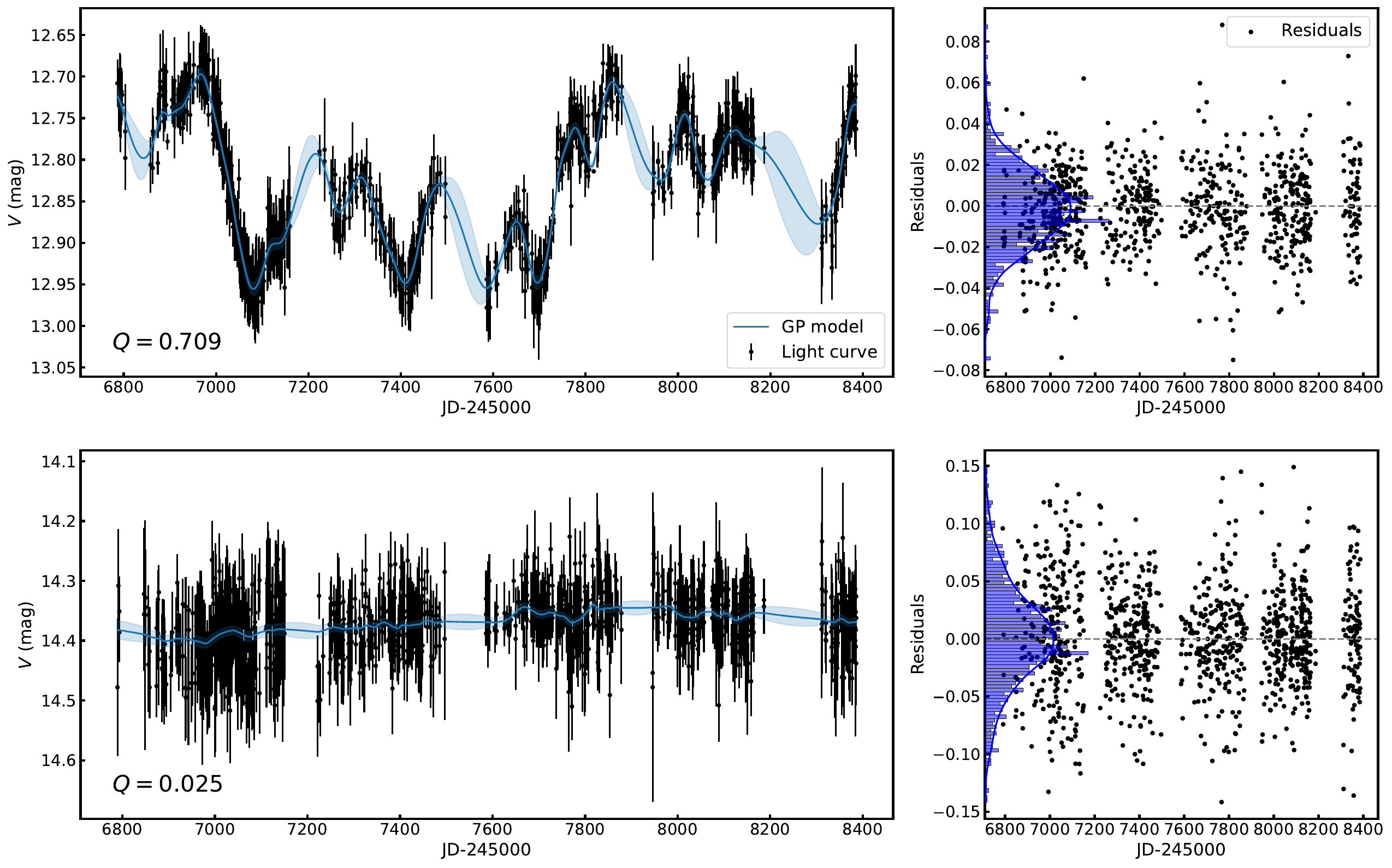}
	\caption{Examples of two $V$-band light curves from the LMC (IDs: top: 10024, bottom: 4635). The upper left and lower left panels display light curves with $Q$ values situated within the right and left Gaussian profiles in Figure \ref{fig: Q_distribution}, respectively, with $Q=0.709$ for the upper left panel and $Q=0.025$ for the lower left panel. The upper right and lower right panels show the residuals of the light curves fittings, defined as the observed magnitudes minus the model magnitudes. The distribution of these residuals and their kernel density estimation are depicted in blue in the figure.}\label{fig: light_curve}
\end{figure}

The inferred values of $Q$ for each band are presented in the histograms in Figure \ref{fig: Q_distribution}. On a logarithmic scale, $Q$ exhibits a bimodal distribution: one group is centered around $\log Q = -1.5$, which is near $Q = 0.03$, while the other is concentrated around $\log Q = 0$, approximately $Q \sim 1$. We select one light curve of these two groups in the $V$-band to examine their variability characteristics, as shown in Figure \ref{fig: light_curve}. Obviously, the GP regression modeling of the light curve in the lower left panel is inadequate, capturing almost none of the variation components, fitting only the long-term trend; in contrast, the fitting in the upper left panel is successful, as it clearly depicts the irregular variability components within the light curve. Further examinations indicates that the light curves with smaller $Q$ values are almost faint RSGs, while larger $Q$ values correspond to brighter RSGs. This is understandable because the granulation in fainter sources is considered to have smaller amplitudes and shorter timescales (see Section \ref{sec: scaling_relations} and Figures 9 and 10 of \citet{2020ApJ...898...24R}), which are difficult to detect with ground-based telescopes due to non-uniform and overly long cadence, thus requiring higher quality data for investigation. On the other hand, as noted in Section \ref{sec: GP_regression}, a PSD with $Q = 1/\sqrt{2}$ is considered to describe granulation signals, and \citet{2022A&A...668A.134B} also found the $Q$ values for their 30 OB stars to span from 0.18 to 2.15, clustering around approximately $1/\sqrt{2}$. Therefore, it is believed that the peak on the right side of Figure \ref{fig: Q_distribution} (corresponding to larger $Q$ values) indeed represents the expected stochastic variations.

A kernel density estimation (KDE) is performed on the bimodal distribution shown in Figure \ref{fig: Q_distribution}. Subsequently, a double-Gaussian fitting is applied to the KDE to characterize the parameters of each peak. The trough value of the double-Gaussian fitting is chosen as the lower boundary of the range for $Q$, and the symmetric value concerning the mean of the right Gaussian profile is selected as the upper boundary for the range of $Q$. The results for each band are listed in Table \ref{tab: Q_range}. Notably, the $I$-band light curves in the SMC and LMC have smaller $Q$ values, corresponding to fainter sources, because bright RSGs tend to saturate in OGLE $I$-band photometry. This makes it challenging to identify the right Gaussian profile in the $I$ band. Indeed, the double Gaussian fitting is not successful for the LMC. Therefore, the range of $Q$ selected for the SMC is likewise applied to the LMC. 

\begin{figure}[h]
	\centering
    \includegraphics[width=0.5\textwidth]{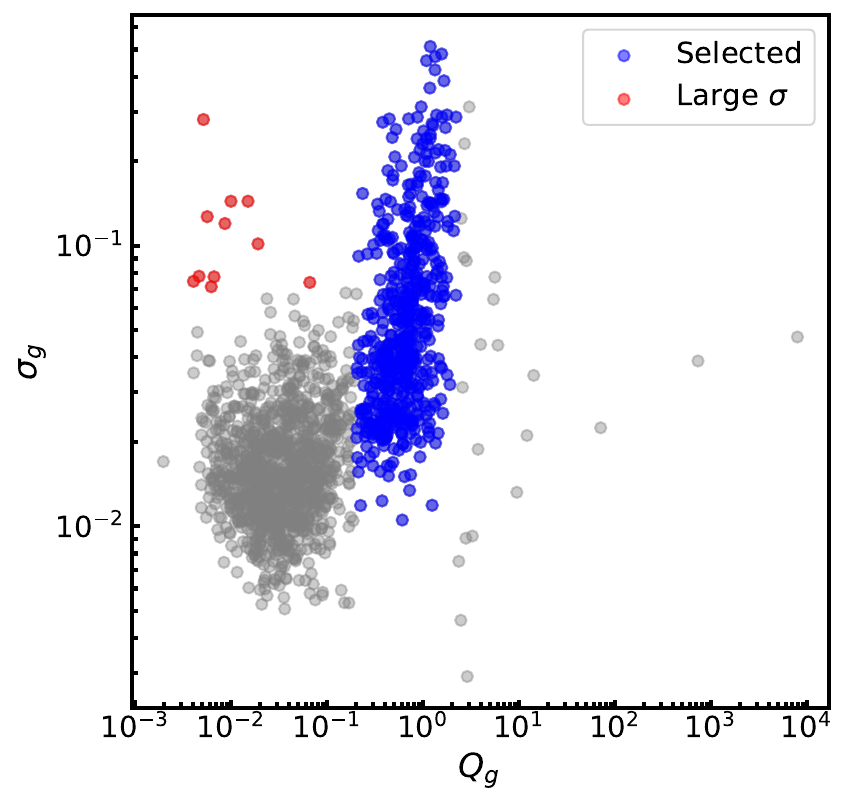}
	\caption{The comparison of $Q$ versus $\sigma$ for the $g$-band light curves of the LMC RSGs. Blue and red points denote sources selected within the bimodal distribution of $Q$ and those with small $Q$ but large $\sigma$, respectively.}\label{fig: Q_sigma}
\end{figure}

Figure \ref{fig: Q_sigma} shows the results of $Q$ versus $\sigma$ for the $g$-band light curves of the LMC RSGs. Selected sources and those discarded are found to separate into two groups: the former spanning a wide range of $\sigma$, while the latter clustering at lower $\sigma$ values. This can be comprehended as smaller $Q$ implying a damping timescale of the oscillators significantly shorter than the undamped period, whereby any deviation from the mean is immediately suppressed, resulting in smaller $\sigma$ values. Light curves with small $Q$ but large $\sigma$ may also possess stochastic signals; however, there are very few of them, i.e., 11, 2, and 7 in the $g$, $V$, and $I$ bands of the LMC RSGs, respectively, and 3, 1, and 4 in the $g$, $V$, and $I$ bands of the SMC RSGs, respectively\footnote{For sources with $Q<Q_{\text{min}}$, Gaussian fitting is applied to the $\sigma$ distribution, where $Q_{\text{min}}$ denotes the lower limit of $Q$ in the selection criteria. These numbers denote the count of light curves with relatively larger $\sigma$ beyond the $\mu+3\sigma$ range of the Gaussian fittings.}. Considering these sources do not substantially influence subsequent results, they are not included in further analyses.

Through such selections, the $g$, $V$, and $I$ bands of the SMC are left with 207, 194, and 55 sources, respectively; for the LMC, the $g$, $V$, and $I$ bands are left with 545, 685, and 92 sources, respectively. Their granulation parameters are listed in Table \ref{tab:params_smc} for the SMC and Table \ref{tab:params_lmc} for the LMC.

\subsection{Comparison between different bands} \label{sec: amplitude_ratio}

\begin{figure}[h]
	\centering
    \includegraphics[width=0.5\textwidth]{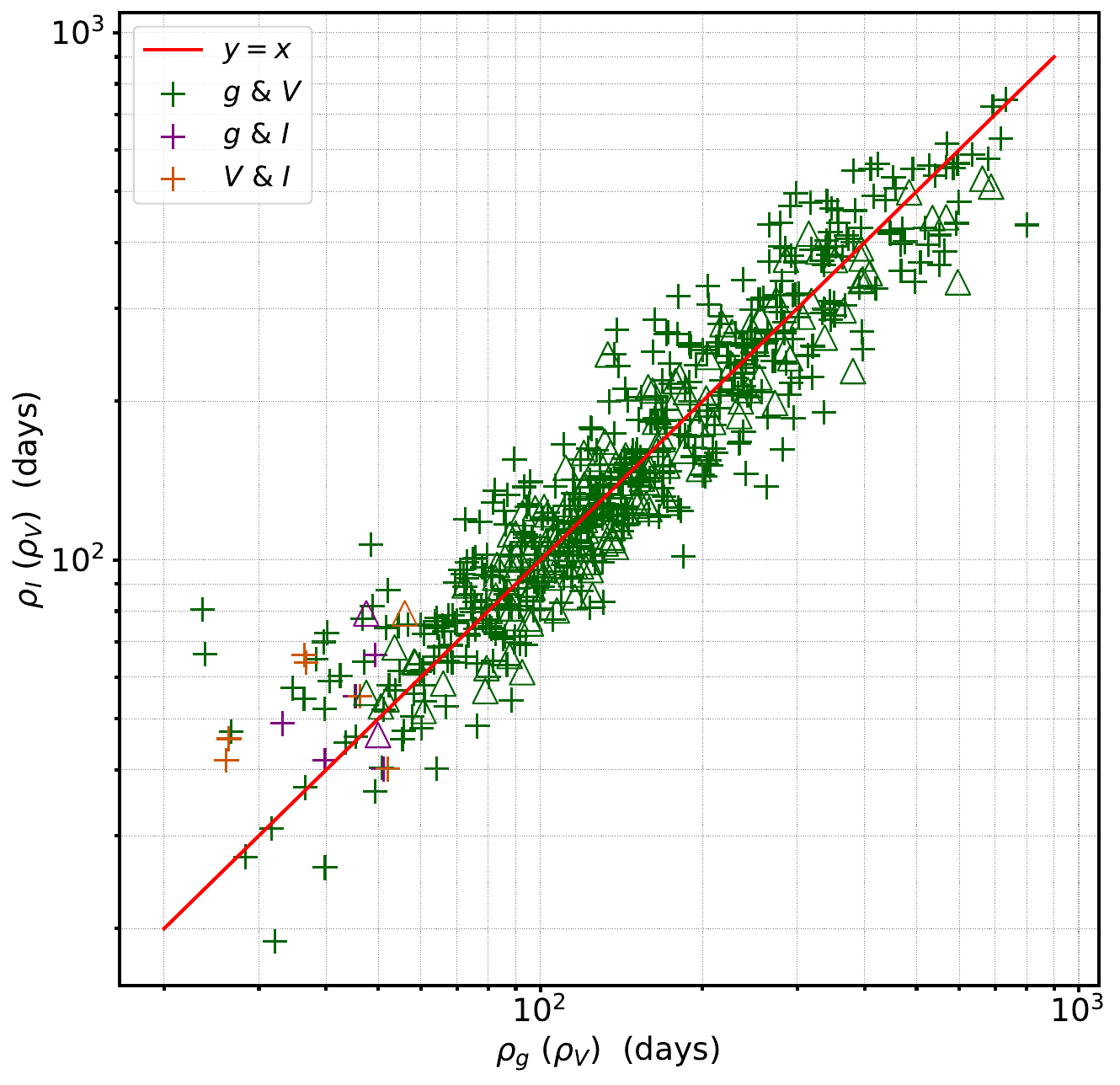}
	\caption{Comparison of granulaiton timescale $\rho$ measured across different bands. Green, purple, and orange points represent comparisons between the $g$ and $V$, $g$ and $I$, and $V$ and $I$ bands, respectively. Triangles and plus signs denote results from the SMC and LMC, respectively. The red solid line marks the one-to-one line.}\label{fig: rho_comparison}
\end{figure}

The determined granulation parameters are used to compare across different bands. Among all the sources remaining in Section \ref{sec: Q_distribution}, for the LMC (SMC), the number of sources obtained by cross-matching two bands is as follows: $g$ and $V$ bands: 469 (145), $V$ and $I$ bands: 6 (1), $g$ and $I$ bands: 5 (2). The substantial difference in numbers can be understood, recalling that the $g$ and $V$ bands come from the same facility and are biased towards brighter sources, whereas the $I$ band provided by OGLE can only observe fainter sources. Figure \ref{fig: rho_comparison} shows the comparison of the granulation timescale $\rho$ with pairings of the three bands distinguished by color, for example, for all green points, the horizontal and vertical axes represent $\rho$ measured in the $g$ and $V$ bands, respectively. The results for the SMC and LMC are denoted by triangles and plus signs, respectively. It appears that almost all points align with one-to-one line, demonstrating the consistency and reliability of light curves measurements across different bands.

\begin{figure}[h]
	\centering
    \includegraphics[width=0.5\textwidth]{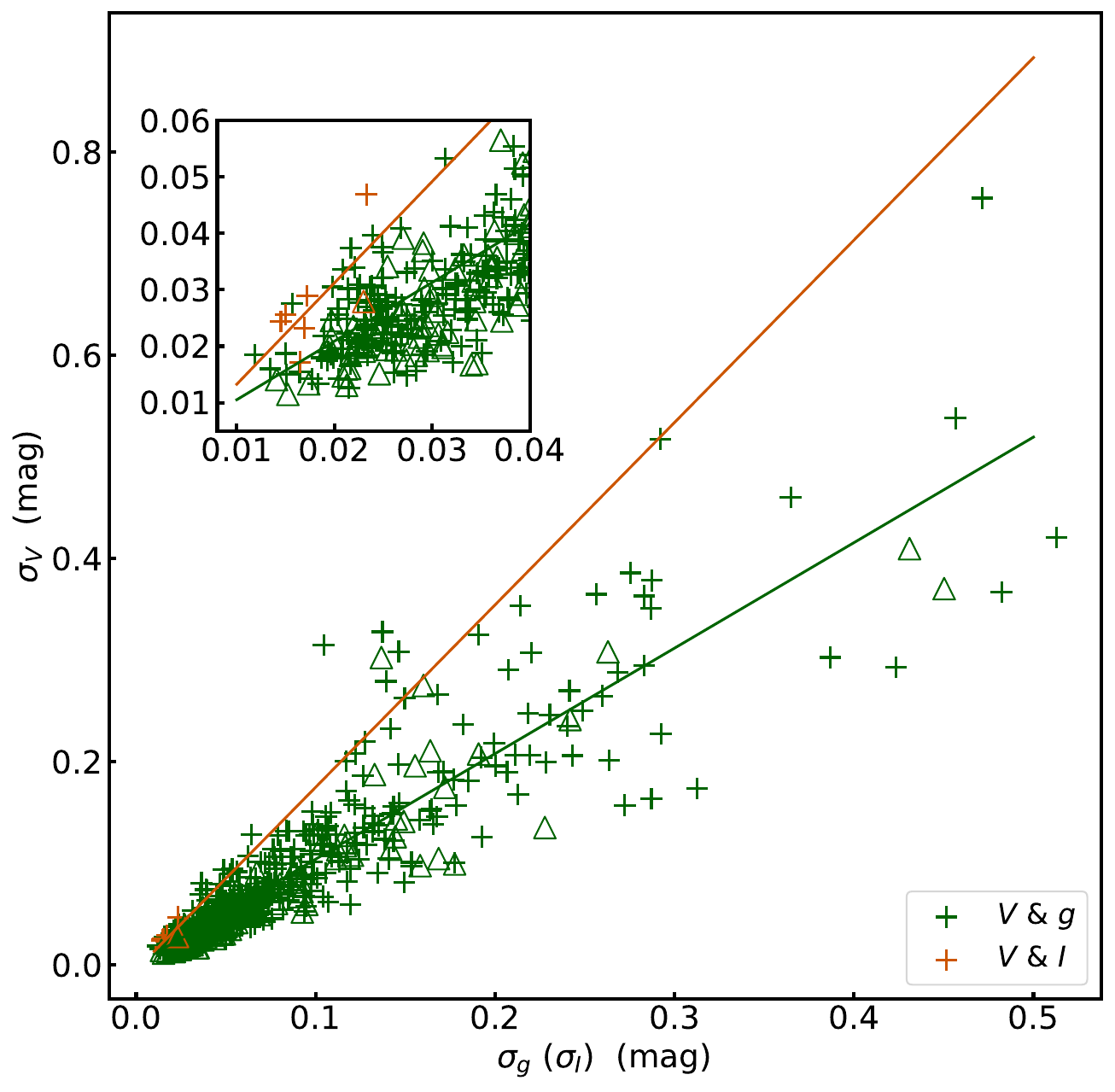}
	\caption{Comparison of granulation amplitude $\sigma$ between the $V$ and $g$ bands, and the $V$ and $I$ bands. The definitions of markers and colors are the same as Figure \ref{fig: rho_comparison}. Orange and green solid lines represent the linear fittings for the two sets of colored points, respectively, used to determine the amplitude ratio. The inset in the upper left corner of the figure highlights the situation around small amplitudes.}\label{fig: sigma_comparison}
\end{figure}

As for the amplitude, the results measured in different bands are expected to differ in the way that optical bands would exhibit larger amplitudes than infrared bands. To determine the amplitude ratios, the granulation amplitudes $\sigma$ measured in the $V$ and $I$ bands, as well as the $V$ and $g$ bands, are presented in Figure \ref{fig: sigma_comparison}. The amplitude ratios between the $V$ and $I$ bands and the $V$ and $g$ bands, $A_{V/I}$ and $A_{V/g}$, are derived to assist in building scaling relations later in the text. By combining the results from the SMC and LMC and performing a linear fitting, $A_{V/I}=1.795$ and $A_{V/g}=1.038$, with correlation coefficients of 0.70 and 0.91, respectively. In previous studies, the amplitude ratio between the $V$ and $R$ bands was considered to be 1.2-1.3 \citep{2019MNRAS.489.1072L} or 1.5 \citep{2009ASPC..412..179P}.

\section{Granulation and stellar parameters} \label{sec: scaling_relation}
\subsection{Stellar parameters} \label{sec:stellar_params}

The granulation parameters are explored for their relation with the essential stellar parameters, namely, effective temperature ($T_{\mathrm{eff}}$), luminosity ($L$), mass ($M$), radius ($R$), and surface gravity ($\log g$).

The 1941 and 4266 RSGs in SMC and LMC selected in this work (described in Section \ref{sec:RSGs_sample}) are searched in the APOGEE/DR17 catalog \citep{2022ApJS..259...35A}, which yields $T_{\mathrm{eff}}$ for 493 and 1475 RSGs respectively. The relation between $T_{\mathrm{eff}}$ of these sources and $(J-K)_0$ is then fitted twice that the points deviating from 95\% confidence interval in the first fit are removed in the second fit to obtain a reliable $T_{\mathrm{eff}} - (J-K)_0$ relations. The intrinsic color index $(J-K)_0$ are adopted from \citet{2021ApJ...923..232R}. The results are as follows:

\begin{equation}
     \frac{T_{\mathrm{eff}}}{1000K}=5.74-1.83(J-K)_0
\end{equation}
for SMC, and
\begin{equation}
     \frac{T_{\mathrm{eff}}}{1000K}=5.42-1.53(J-K)_0
\end{equation}
for LMC.

\begin{figure}[h]
	\centering
    \includegraphics[width=\textwidth]{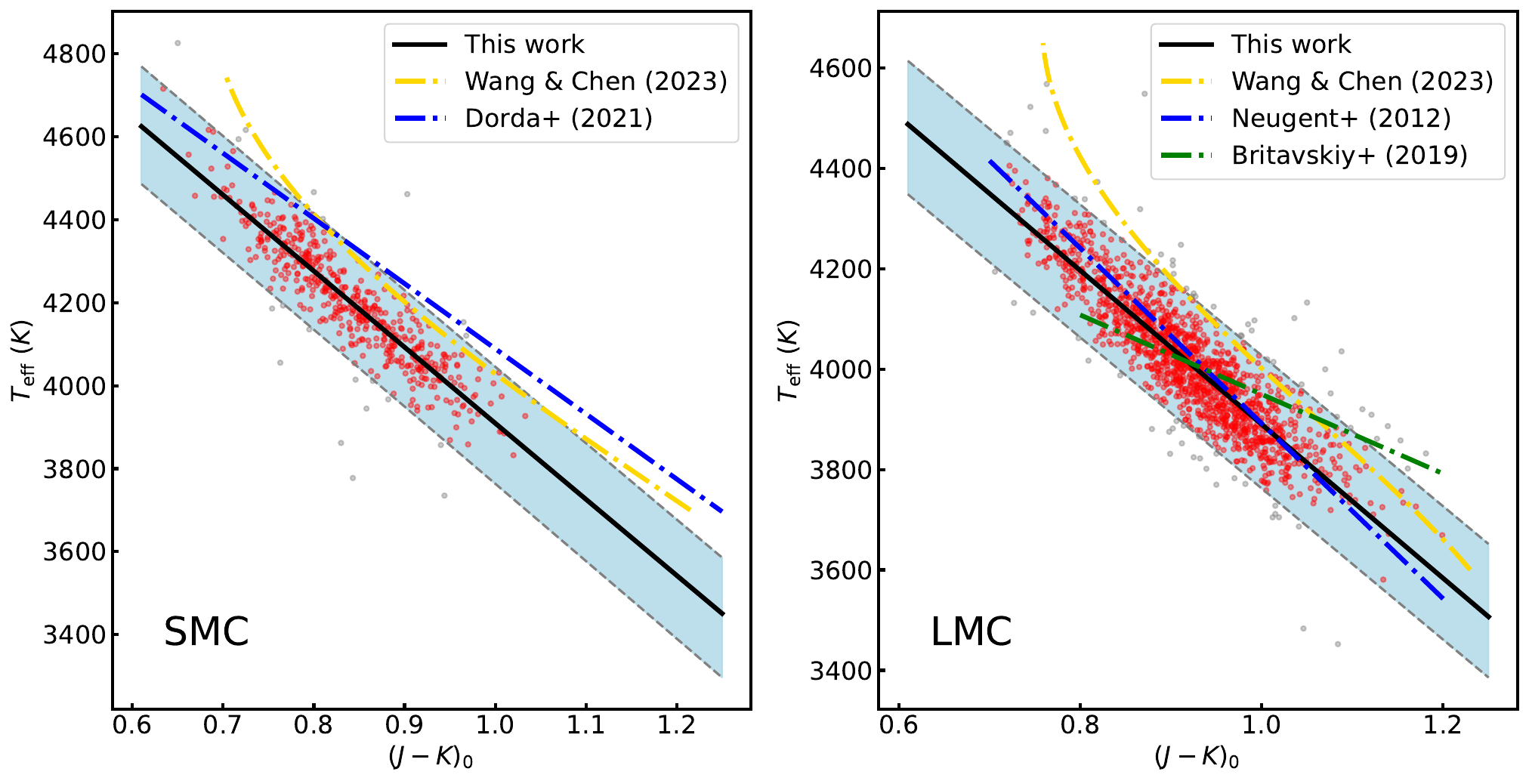}
	\caption{Relations between $T_{\mathrm{eff}}$ and $(J-K)_0$ of RSGs. The light blue shaded area represents the 95\% confidence intercal obtained from the first fitting, while the solid black line illustrates the final fitting result. The red dots represent the sources used in the fitting process, whereas the gray dots lie beyond the 95\% confidence intercal of the first fitting and are excluded from the second fitting. The dashed blue, green, and the gold lines represent results obtained from other studies. \label{fig:teff_color}}
\end{figure}

The results are displayed in Figure \ref{fig:teff_color} and compared with previous relations by \citet{2021MNRAS.502.4890D} and \citet{2023ApJ...946...43W} for the SMC and \citet{2012ApJ...749..177N}, \citet{2019A&A...624A.128B} and \citet{2023ApJ...946...43W} for the LMC. The slope agrees very well with others except \citet{2019A&A...624A.128B}, and the slight systematic shifts may come from different extinction correction. For instance, the intrinsic colors calculated by \citet{2021ApJ...923..232R} are derived from the 2D extinction maps based on red clump stars provided by \citet{2021ApJS..252...23S}, while \citet{2023ApJ...946...43W} built their own extinction map using the color excess of RSGs identified by the APOGEE stellar parameters.

\citet{2013ApJ...767....3D} derived a formula to convert the apparent magnitude $m_{\lambda}$ and distance modulus $\mu$ to luminosity for RSGs in nearby galaxies like following:
\begin{equation}
     \log(L/L_{\odot})=a+b(m_{\lambda}-\mu)
\end{equation}
where the value of $a$ and $b$ is $0.90\pm 0.11$ and $-0.40\pm 0.01$ respectively for the $K$ band. The $K$ band is chosen because RSGs are bright at this band and are hardly affected by interstellar extinction. The distance modulus of SMC and LMC are taken as 18.95 \citep{2016ApJ...816...49S} and 18.49 \citep{2013Natur.495...76P}, respectively. Then, the mass $M$ is calculated from the mass-luminosity relation $L/L_{\odot}=(M/M_{\odot})^\gamma$, and for RSGs, the exponent $\gamma=4$ \citep{1971A&A....10..290S}. With $L$, $T_{\mathrm{eff}}$ and $M$, $R$ and $g$ can be calculated from $L=4\pi R^2 \sigma T^4$ and $g=GM/R^2$, respectively.

\begin{figure}[h]
	\centering
    \includegraphics[width=\textwidth]{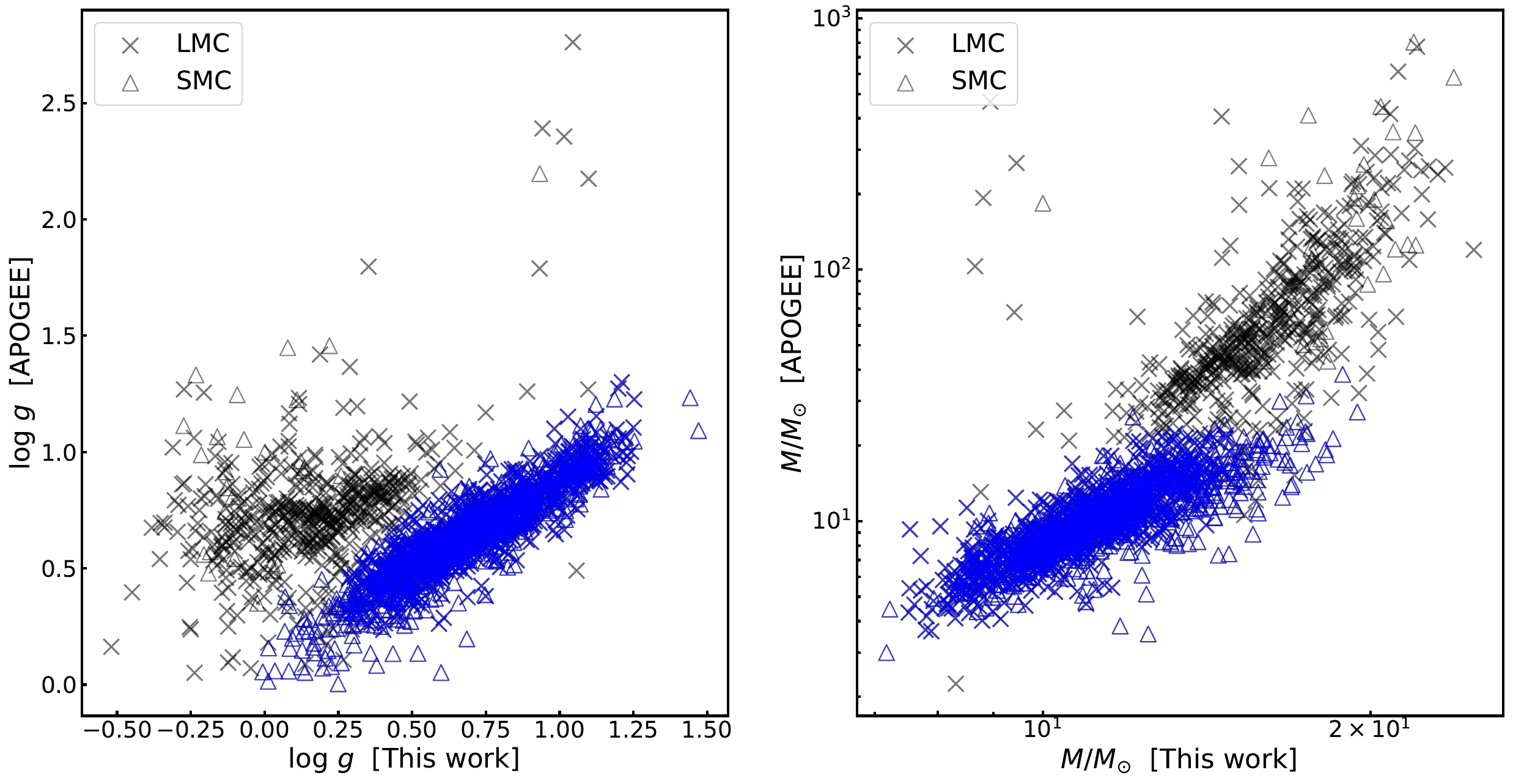}
	\caption{The comparison of surface gravity (left panal) and stellar mass (right panel) between the APOGEE measurement and this work. The sources of the LMC and SMC are marked with crosses and triangles, respectively. Sources with consistent $\log\ g$ measurements are marked in blue, while others are represented in black.} \label{fig: SP_comparison}
\end{figure}

The spectroscopic stellar parameters from APOGEE are compared with the aforementioned results, as shown in Figure \ref{fig: SP_comparison}. The left panel shows the comparison for $\log\ g$, where most sources shows good consistency, but a few deviate significantly. This could be due to systematic bias of APOGEE, as a similar trend is also found in Figure 1 of \citet{2023ApJ...946...43W}. These sources also exhibit large APOGEE fitting chi-squares, indicating that the inferred parameters might not be as accurate as others. Because these sources are at the brightest end of the RSGs, the model grids may be unreliable under such low surface gravity conditions. The right panel illustrates the comparison of stellar masses, where $M/M_{\odot}\ [\mathrm{APOGEE}]$ is calculated using $\log \ g$ from APOGEE and radius $R$ from $L/T_{\mathrm{eff}}$. RSGs with consistent $\log \ g$ measurements are marked in blue, including 1011 sources in the LMC and 461 in the SMC. They also show consistency in mass comparison, which is reasonable as the difference in mass calculation only comes from surface gravity, thus $\log \ g$ dominates the outliers in the right panel. Considering RSGs shouldn't physically have masses larger than about $40 M_{\odot}$, this further proves the inaccuracy of APOGEE's surface gravity measurements for these stars, while our calculation is reliable.

\subsection{The scaling relations} \label{sec: scaling_relations}

\subsubsection{Relation between $\rho$ and stellar parameters} \label{sec: scalingrelation_rho}

The relations between granulation parameters and stellar parameters are considered both observationally and theoretically, which are mainly derived under some assumptions and hydrodynamic simulations. \citet{2009CoAst.160...74H} and \citet{2011Natur.471..608B} show that the timescale of granulation $\rho$ is proportional to the pressure scale height $H_p$, and inversely proportional to the sound speed $c_\mathrm{s}$, that is, $\rho \propto H_p/c_\mathrm{s}$. Because of $H_p\propto T_{\mathrm{eff}}/g$ and $c_\mathrm{s} \propto \sqrt{T_{\mathrm{eff}}}$ \citep{1995A&A...293...87K}, $\rho \propto \sqrt{T_{\mathrm{eff}}}/g$.

\begin{figure}[h]
     \centering
     \includegraphics[width=\textwidth]{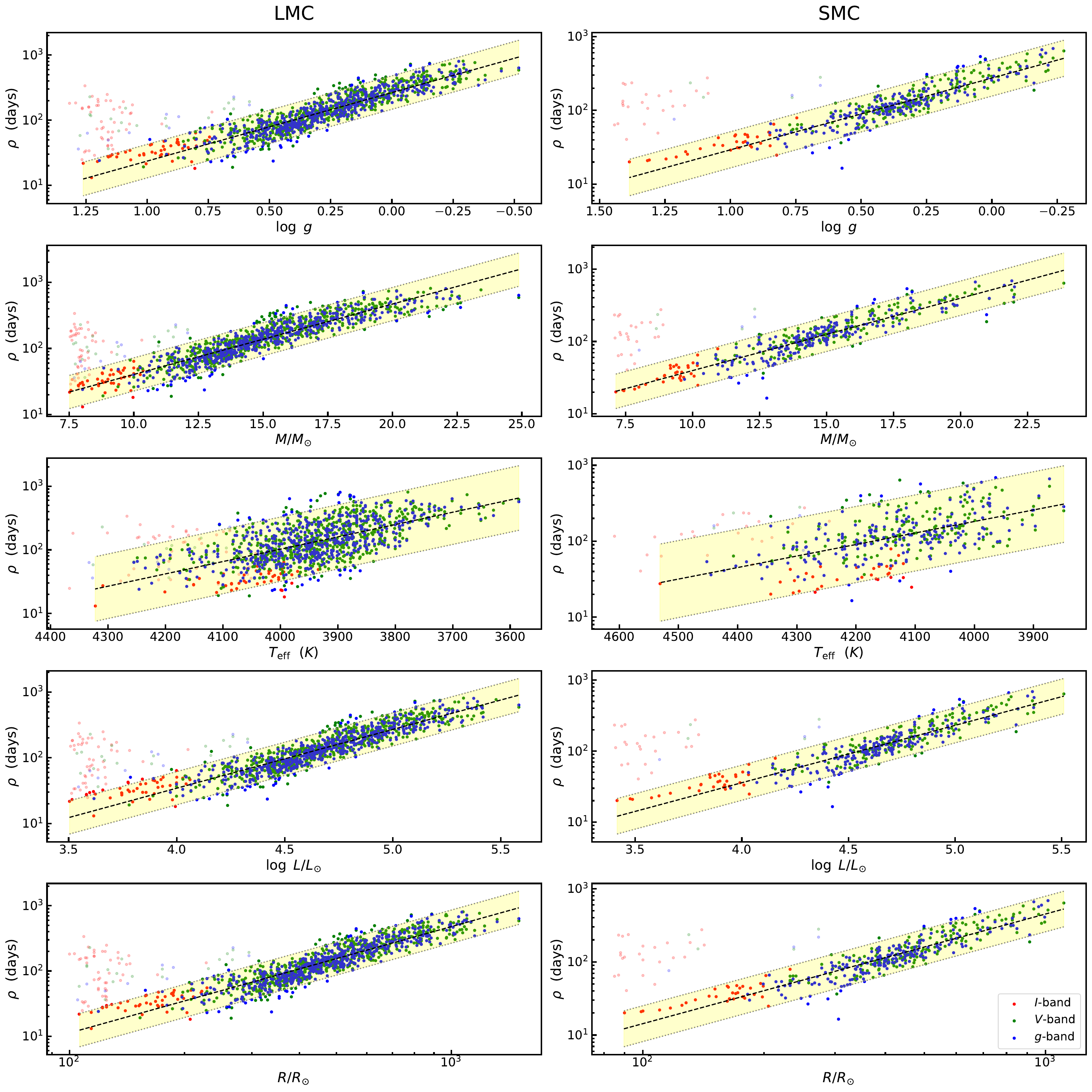}
	\caption{Scaling relations of granulation timescale $\rho$ with stellar surface gravity, mass, effective temperature, luminosity, and radius (from top to bottom) for RSGs in the Magellanic Clouds. The results of LMC and SMC are shown in the left and right column, respectively. The red, blue, green dots shows results from $I$, $g$ and $V$ bands, respectively. The black dashed lines show the best fitting of the scaling relations with the light yellow shadow covering the 95\% confidence interval. The outliers are transparently displayed in this figure. \label{fig: scalingrelation_rho}}
\end{figure}

The relations between $\rho$ and stellar parameters of RSGs in the LMC and SMC are displayed in Figure \ref{fig: scalingrelation_rho}, demonstrating a significant linear correlation. Some points, particularly in the faint end of the $I$ band, exhibit larger $\rho$ values, leading to their apparent deviation from the scaling relations. Indeed, these sources possess larger damping timescales ($\tau$), exhibiting less stochastic. Upon examining light curves, a subset of these sources are found to exhibite quasi-periodic low-frequency variability, with their PSDs showing a peak at low frequencies corresponding to the determined $\rho$ values. Therefore, it is possible that the sample might be contaminated by stars of unknown types, with other mechanisms driving their variability. For these reasons, these points are manually excluded when fitting the scaling relations. Their analytical expressions are listed in Table \ref{tab: rho_scaling_relations}.

\subsubsection{Relation between $\sigma$ and stellar parameters}

\citet{2011A&A...529L...8K} indicates that the fluctuations arise from a large number of granules on stellar surface, so the rms of these fluctuations (i.e., $\sigma$) is thought to scale inversely with the square root of the number of granules $n$, and is proportional to the sound speed $c_\mathrm{s}$. Since the diameter of the granules is assumed to be proportional to the pressure scale height of the atmosphere \citep{1975ApJ...195..137S},  $n\propto({R}/{H_p})^2$. Thus, $\sigma \propto T_{\mathrm{eff}}^{1.5}/gR$.

\begin{figure}[h]
     \centering
     \includegraphics[width=\textwidth]{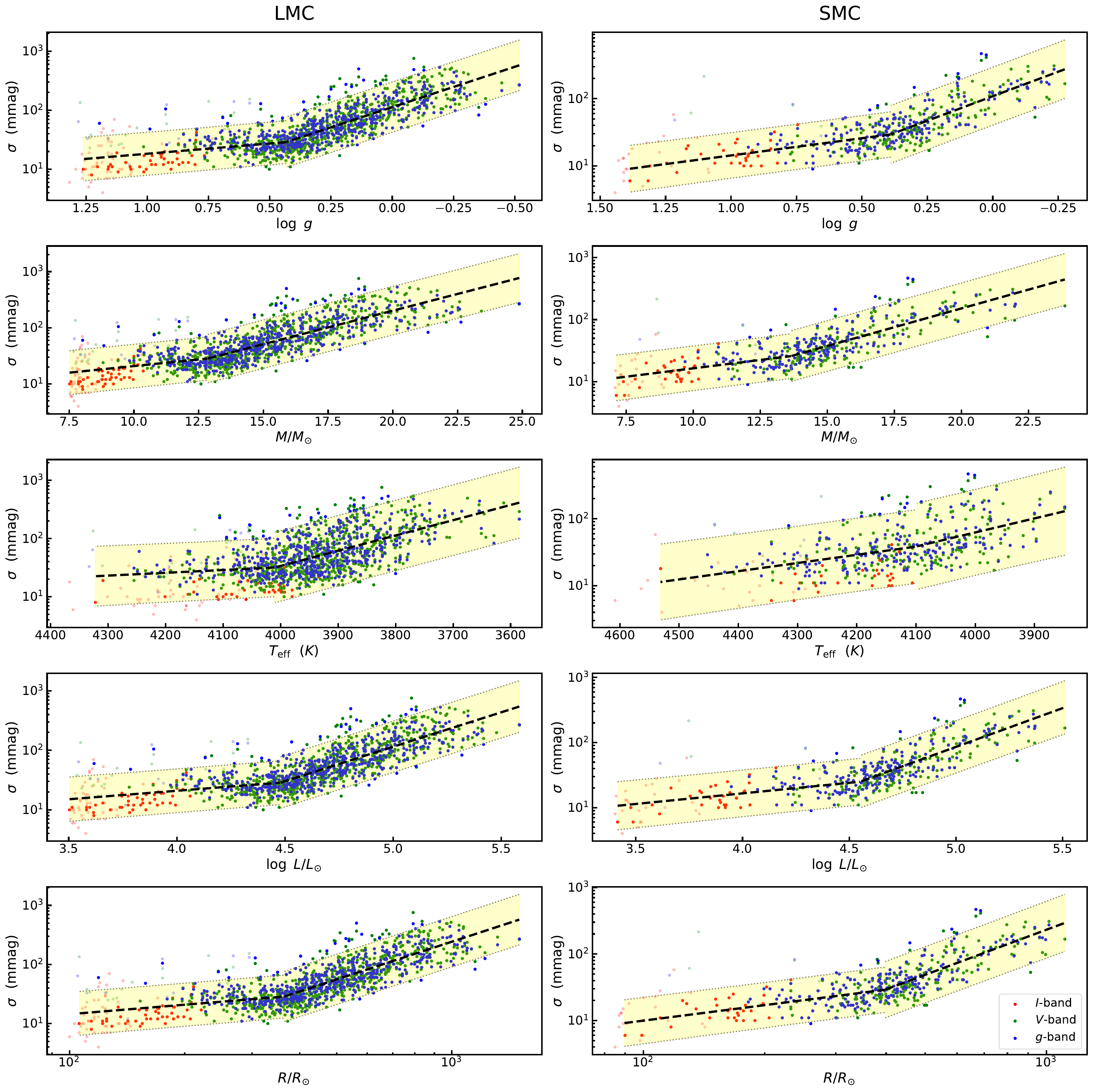}
	\caption{Same as Figure \ref{fig: scalingrelation_rho}, but for $\sigma$. The fitting are performed using piecewise functions.\label{fig: scalingrelation_sigma}}
\end{figure}

Figure \ref{fig: scalingrelation_sigma} illustrates the relations between $\sigma$ and stellar parameters in the SMC and LMC. Data from the $g$ and $I$ bands are multiplied by 1.038 and 1.795, respectively, which are the amplitude ratios obtained in Section \ref{sec: amplitude_ratio}. Points that are not used for fitting the scaling relations of $\rho$ (Section \ref{sec: scalingrelation_rho}) are also depicted in a transparent manner. However, it appears that RSGs at the bright and faint ends follow different relations; specifically, the bright end exhibits a steeper slope, but this relation does not extend to the faint end and tends to flatten after a certain inflection point. This relation between amplitudes and stellar parameters is also identified in other RSGs samples (e.g., Figure 4 of \citet{2018ApJ...859...73S}, Figure 9 of \citet{2019MNRAS.487.4832C}, and Figure 12 of \citet{2023A&A...676A..84Y}), showing high consistency with this work. This phenomenon is thought to indicate the presence of some ``non-varying" RSGs within the RSG population, which are generally characterized by low luminosity, also leading to the piecewise structure in the relation between luminosity and mass loss rate \citep{2023A&A...676A..84Y, 2024AJ....167...51W}. Therefore, we use a piecewise function to fit the scaling relations of $\sigma$, and its mathematical expression is provided in Table \ref{tab: sigma_scaling_relations}.

\section{Discussion} \label{sec: discussion}

\subsection{Metallicity effects}

Comparing the granulation parameters between the two galaxies is a natural consideration that can reveal the characteristics of granulation under different environments. We calculate the median values of the granulation parameters described in Section \ref{sec: scaling_relations}. In the $V$ band, the median $\sigma$ for the LMC is 47.8 mmag, compared to 37.4 mmag for the SMC; in the $g$ band, the median $\sigma$ values for the LMC and SMC are 47.6 mmag and 35.4 mmag, respectively. This indicates that, on average, the LMC exhibits larger granulation amplitudes than the SMC. The results for the timescale $\rho$ demonstrate a similar trend, with median values in the $V$ band being 137.3 days for the LMC and 130.0 days for the SMC; in the $g$ band, the median values are 132.0 days for the LMC and 113.5 days for the SMC. The $I$-band data are excluded due to their limited number and significant bias towards faint end. One obvious reason of the above trends could be the metallicity effect, as the two galaxies have very different metallicity, with the LMC being about twice as large as the SMC. The effect of metallicity on granulation was studied by numerical simulation \citep{2013A&A...560A...8M} or observation \citep{2017A&A...605A...3C, 2018ApJS..236...42Y}. In recent studies, \citet{2017A&A...605A...3C} found that metallicity leads to significant variations in granulation amplitudes and timescales of RGBs in three open clusters, NGC 6791, NGC 6819, and NGC 6811, which span the metallicity range [Fe/H] from $-$0.09 to 0.32. \citet{2017A&A...605A...3C} pointed out that, an increase in metallicity leads to an increase in opacity and mixing length, giving granulation a larger size and therefore an increase in amplitude and timescale. This was also confirmed by \citet{2018ApJS..236...42Y}, who found that metal-rich RGBs have larger granulation power (described as $\sigma_{\mathrm{gran}}^2\tau_{\mathrm{gran}}$) than metal-poor stars. For RSGs, \citet{2020ApJ...898...24R} drew a similar conclusion, namely that the amplitude and timescale of granulation increase with metallicity.

\begin{figure}[h]
     \centering
     \includegraphics[width=\textwidth]{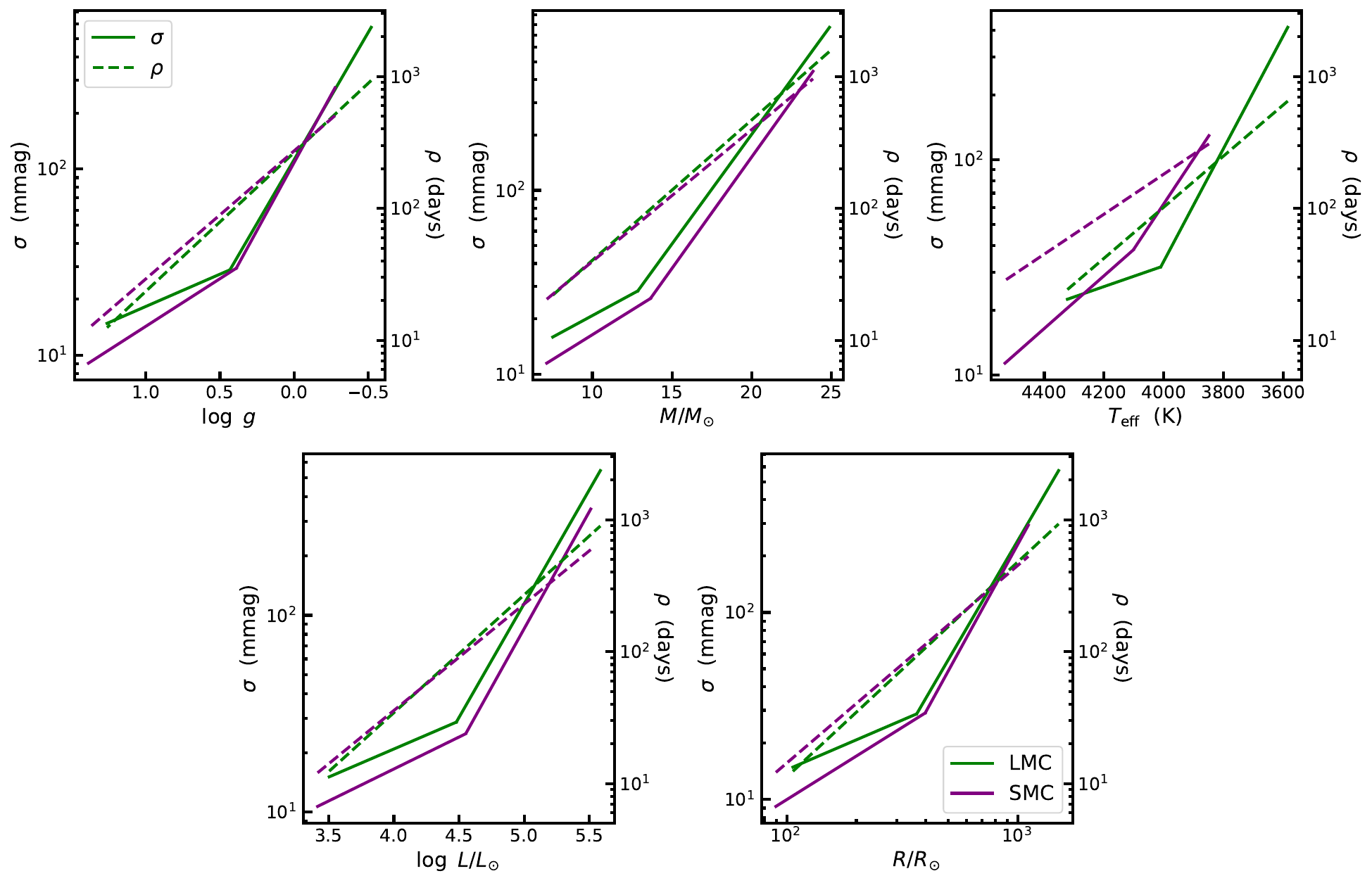}
	\caption{Scaling relations in the Magellanic Clouds. The green and purple lines are the results of LMC and SMC, respectively. The solid and dashed lines are scaling relations of $\sigma$ and $\rho$, which is expressed by the left and right Y axes, respectively.\label{fig: SR_MCs}}
\end{figure}

The scaling relations are compared between SMC and LMC in Figure \ref{fig: SR_MCs} for $\sigma$ and $\rho$, respectively, where the horizontal axis range coincides with that of stellar parameters of our RSGs sample. Although the 95\% confidence intervals cover the differences in scaling relations between the two galaxies, some patterns can still be discerned. Most notably, the scaling relations for $\rho$ between the LMC and SMC are remarkably similar, whereas the relations for $\sigma$ exhibit more significant deviations. This observation aligns with the findings of \citet{2020ApJ...898...24R}, who compared the granulation parameters with stellar parameters between the LMC RSGs and Kepler RGBs. They discovered that the timescales $\tau_{\mathrm{eff}}$ for RSGs generally coincide with the extrapolation of the relations for RGBs (see their Figure 9), while the amplitude $\sigma_{\mathrm{gran}}$ did not exhibit any discernible relations across the two stellar populations (see their Figure 10). This suggests that granulation timescales may not be sensitive to metallicity, whereas amplitudes are more affected by metallicity. This implies that the impact of chemical composition of the stellar atmosphere on the convective envelope is complex and requires further simulations through more precise modeling and confirmation with a larger sample.

On the contrary, $T_{\mathrm{eff}}$ exhibits an quite different tendency, i.e.  $\sigma$ and $\rho$ in the SMC is larger than in the LMC at a given $T_{\mathrm{eff}}$. This may be due to the difference in $T_{\mathrm{eff}}$ among RSGs in the two galaxies compared to other stellar parameters, with $T_{\mathrm{eff}}$ in the SMC larger that in the LMC, causing an overall shift of the scaling relation towards the high-temperature end. In other words, the effect of metallicity on $T_{\mathrm{eff}}$ is more significant than on granulation.

\subsection{Comparison of GP regression and PSD fitting}

\begin{figure}[h]
     \centering
     \includegraphics[width=0.6\textwidth]{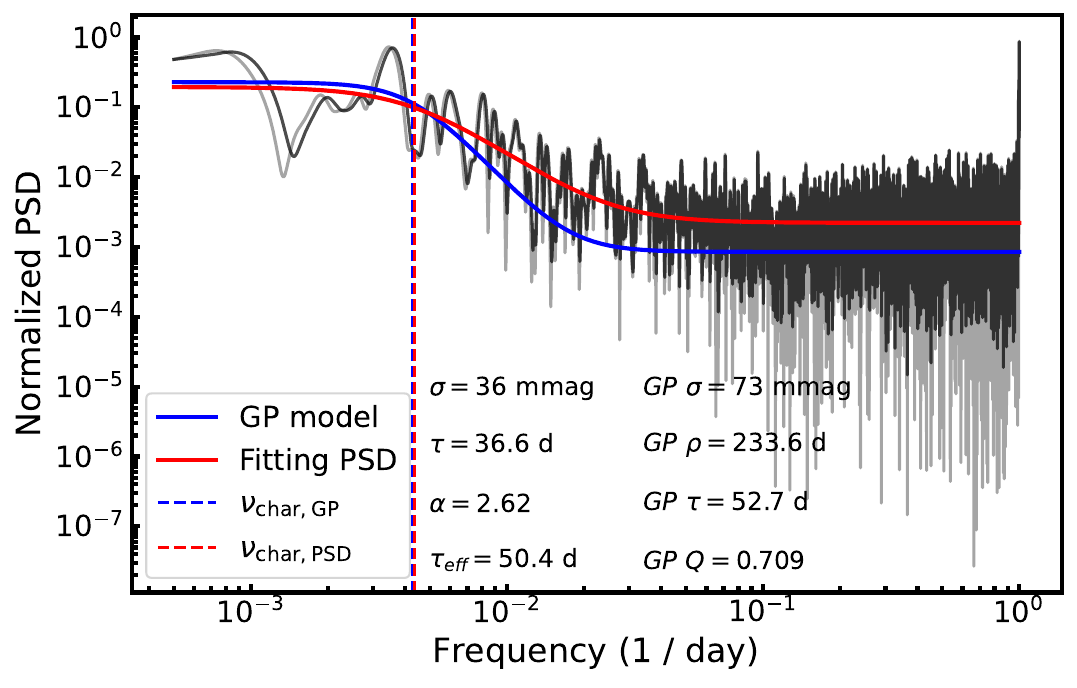}
	\caption{Example of PSD fitting for the ASAS-SN $V$-band light curve (ID: 10024). The PSD of this light curve is shown in gray, while its smoothed version is presented in black. The blue and red solid lines show the granulation signal modeled by the GP regression and PSD fitting, respectively. The blue and red dashed lines mark the position of $\nu_{\mathrm{char}}$ for each method. The fitting parameters are displayed in the figure. \label{fig: psd}}
\end{figure}

GP regression modeling of light curves exhibit strong flexibility and robustness. For comparison, the traditional method of fitting PSD is also applied to these light curves. While fitting PSD may not be sufficiently accurate due to numerous potential biases, understanding the strengths and applicability of both methods is meaningful. The PSD is calculated using the Lomb-Scargle algorithm which is built in {\sc astropy} \citep{2018AJ....156..123A}, and normalized utilizing Parseval's theorem as described in \citet{2019MNRAS.489.5764P}. Equation \ref{equ: granulation} is used for the PSD fitting, incorporating a white noise term $W$ to account for the high-frequency noise, and replace the exponent with $\alpha$. All fittings are performed in the logarithmic space. Thus, Equation \ref{equ: granulation} is re-written as:
\begin{equation}
     y_\mathrm{model} = \log_{10}\left(W + \frac{4\sigma_{\mathrm{gran}}^2 \tau_{\mathrm{gran}}}{1+(2\pi \nu \tau_{\mathrm{gran}})^\alpha}\right),
\end{equation}
and the likelihood function used for parameter estimation is simplified as:
\begin{equation}
     \ln L = -\frac{1}{2}\sum_{i} ( y_i - y_{\mathrm{model},i} )^2.
\end{equation}
Sampling of the posterior distribution of parameters is conducted using MCMC, with the median of the samples taken as the best-fit parameters. The effective timescale $\tau_{\mathrm{eff}}$ suggested by \citet{2011ApJ...741..119M}, namely $e$-folding time of ACF is adopted to compare the granulation timescale obtained by different $\alpha$. An example of PSD fitting is presented in Figure \ref{fig: psd}, where the PSD from GP regression is displayed for visual comparison, despite GP regression not inherently involving PSD. Figure \ref{fig: psd} also shows the characteristic frequencies $\nu_{\mathrm{char}}$ derived from both methods; for GP regression, defined as $\nu_{\mathrm{char}} = 1/ \rho$, and for PSD fitting, defined as $\nu_{\mathrm{char}} = 1/ (2\pi \tau_{\mathrm{gran}})$.

\begin{figure}[h]
     \centering
     \includegraphics[width=\textwidth]{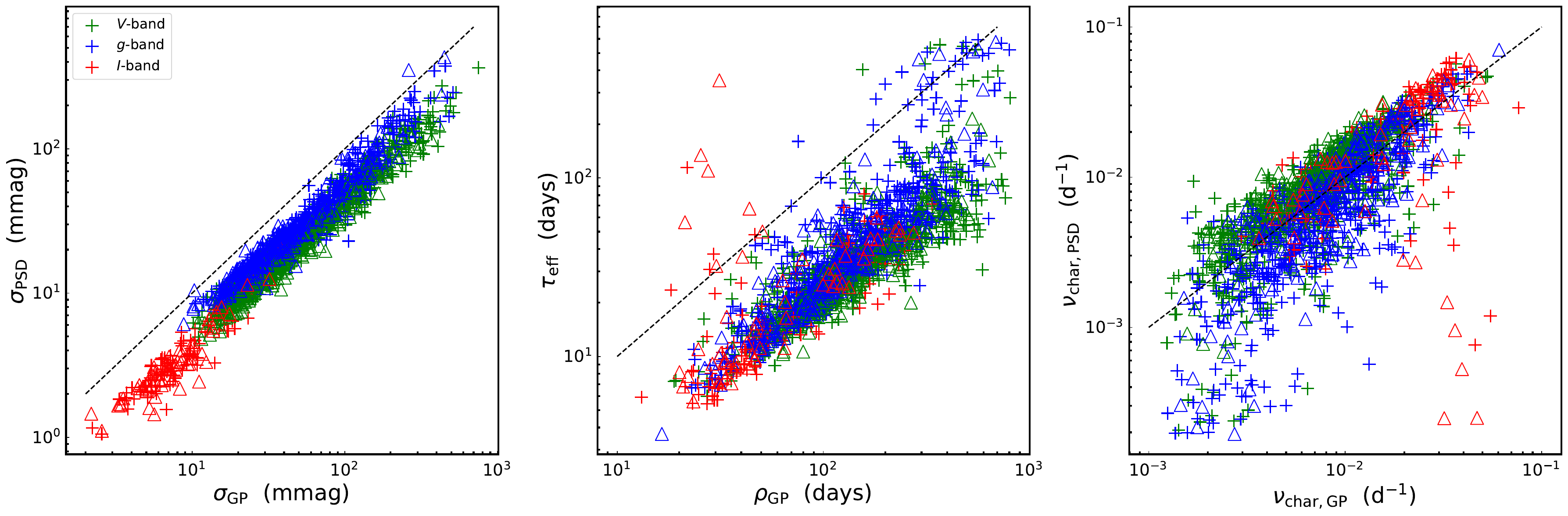}
	\caption{Comparison of parameters derived from GP regression and PSD fitting. The panels from left to right display the results for granulation amplitude, timescale, and characteristic frequency, respectively. Green, blue, and red indicate data from the $V$, $g$, and $I$ bands, with triangles and plus signs representing results from the SMC and LMC, respectively. The black dashed line marks the one-to-one line.\label{fig: GP_PSD_comparison}}
\end{figure}

Figure \ref{fig: GP_PSD_comparison} provides a comparative analysis of the parameters obtained by both GP regression and PSD fitting, with results from the former plotted on the horizontal axis and the latter on the vertical axis. The left panel compares the granulation amplitudes, where the results of both methods show substantial agreement. This consistency is expected as the determination of $\sigma_{\mathrm{gran}}$ relies on the ``plateau" of the PSD at low frequency, which tends to be more straightforward to capture (see Figure \ref{fig: psd}). The middle panel contrasts $\rho$ and $\tau_{\mathrm{eff}}$, revealing a generally good correlation for the majority of sources, albeit with a slightly higher dispersion than amplitude. This dispersion arises because $\tau_{\mathrm{gran}}$ corresponds to the ``knee" of the Lorentzian profile, and the shape of the PSD is greatly influenced by the quality of the time series data, resulting in larger measurement uncertainties. A small subset of sources has $\tau_{\mathrm{eff}}$ significantly larger than $\rho$ due to a failure in correctly identifying the ``plateau" in the PSD, instead exhibiting an upward trend, suggesting a lower $\nu_{\mathrm{char}}$ and hence an overestimated $\tau_{\mathrm{gran}}$. The right panel supplements the middle panel, showcasing the comparison of the characteristic frequency $\nu_{\mathrm{char}}$, which closely aligns with the one-to-one line. Therefore, the advantage of PSD fitting may lie only in amplitude measurement, without showing significant superiority over GP regression. Another potential advantage is its ability to measure the slope of the PSD, $\alpha$, however, more high-quality light curves are required to reconstruct accurate PSDs \citep{2022A&A...668A.134B}, and no clear physical trends are evident with the sample in this work.

\section{Summary} \label{sec:summary}

This work uses the most complete RSGs sample to date, along with the light curves obtained from OGLE in the $I$-band and ASAS-SN in the $g$- and $V$-band, to systematically investigate the granulation characteristics of RSGs in the SMC and LMC. The initial RSG sample undergoes through further removal of foreground stars and potentially misidentified objects, resulting in 1941 and 4266 RSGs in SMC and LMC, respectively, for granulation analysis. Outliers in the light curves are removed, and those exhibiting characteristics of white noise are also excluded. The mean magnitudes of light curves are compared to the {\em Gaia} measurement, which found some sources apparently brighter than the {\em Gaia} measurements. These sources are found to be located in dense star fields in the images, so that their photometry is unreliable and discarded. 

Gaussian Process regression based on {\sc celerite2} and a stochastically driven damped harmonic oscillator kernel is applied for modeling granulation. We select three parameters: $\sigma$, $\rho$, and $\tau$, representing the granulation amplitude, timescale, and the damping timescale of harmonic oscillations, respectively. The dimensionless quality factor $Q$ is calculated through $Q = \pi \tau/\rho$, serving to characterize the features of the variability. The MCMC method is employed to explore the posterior distribution of the parameter space, with the Gelman-Rubin diagnostic ensuring the convergence of the fittings, that is, $\hat{R} < 1.01$.

The parameter $Q$ shows a bimodal distribution across almost all band data: one group centered around $\log Q = -1.5$, and another around $\log Q = 0$. Sources with larger $Q$ values are considered to have a significant granulation signal, while the light curves with smaller $Q$ exhibit almost no noticeable variations, with GP regression modeling only long-term trends. Double-Gaussian fittings are used to characterize the profile of the peak with larger $Q$ values, which is then utilized for further analysis. Parameters determined across different bands are compared, with timescale measurements showing good consistency. The amplitude ratios between different bands are also determined: $A_{V/I}=1.795$ and $A_{V/g}=1.038$.

The scaling relations between granulation parameters and stellar parameters are analyzed, with the timescale $\rho$ exhibiting a clear linear relationship that is remarkably similar between the SMC and LMC. The amplitude $\sigma$ appears to follow different trends at the bright and faint ends, potentially due to the presence of some nearly invariant RSGs causing a flatter slope in the scaling relations at the faint end. Moreover, differences exist in the scaling relations for $\sigma$ between the SMC and LMC. This could be due to the impact of metallicity on amplitude and timescale, with timescale being independent of metallicity, while amplitude is more significantly affected, deserving further investigation. On the other hand, the overall larger granulation amplitude and timescale in the LMC compared to the SMC suggest that a richer metal environment may increase both amplitude and timescale, consistent with findings in the literatures.

The PSD fitting method is compared with GP regression, and it is found that the parameters inferred by both methods generally exhibit consistency. The measurement of amplitude for PSD fitting is relatively accurate but does not demonstrate a significant advantage over GP regression. The determination of timescale is more heavily influenced by the shape of the PSD, resulting in greater dispersion. Therefore, we conclude that GP regression remains a robust method for studying the granulation signal in RSGs, while PSD fitting requires further investigation with higher quality light curves.

\section*{Acknowledgements}
We would like to thank the anonymous referee for the constructive suggestions that definitely improve this work. We thank Dr. Jing Tang and Mr. Haoran Dou for their helpful discussions. This work is supported by the National Natural Science Foundation of China (NSFC) through grants Nos. 12133002 and 12203025, National Key R\&D Program of China No. 2019YFA0405503,  CMS-CSST-2021-A09 and Shandong Provincial Natural Science Foundation through project ZR2022QA064. This work has made use of data from the surveys by OGLE, ASAS-SN, {\em Gaia}, and APOGEE.

%

\vspace{5mm}


\software{astropy \citep{2013A&A...558A..33A,2018AJ....156..123A},
          TOPCAT \citep{2005ASPC..347...29T},
          EMCEE \citep{2013PASP..125..306F},
          LMfit \citep{matt_newville_2018_1699739}}

\bibliography{Modeling_granulation_with_GP}{}
\bibliographystyle{aasjournal}

\clearpage

\begin{deluxetable}{ccccc}[h] \label{tab:RSG_numbers}
	\tablecaption{Number of RSGs removed and retained in our selection process}
	\tablewidth{0pt}
	\tablehead{\colhead{Galaxy} & \colhead{Initial sample} & \multicolumn2c{Removed stars}  & \colhead{The remaining RSGs} \\ \hline
     \colhead{} & \colhead{} & \colhead{By RV} & \colhead{By {\em Gaia} CMD} & \colhead{}}
	\startdata
	SMC & 2,138 & 44$^{a}$ & 153 & 1941\\
	LMC & 4,823 & 133$^{b}$ & 424 & 4266 \\
	\enddata
\tablecomments{$^{a}$The 44 stars are all removed by RV from Gaia. \\
				$^{b}$Of these, 123 and 10 stars are removed by RV from Gaia and APOGEE, respectively.}
\end{deluxetable}

\begin{deluxetable}{cccccc}[h]
     \tablecaption{Number of light curves removed and retained in our selection process \label{tab: LC_number}}
     \tablewidth{0pt}
     \tablehead{
          \colhead{Galaxy}  & \colhead{Band}& \colhead{Count}&\multicolumn2c{Removed light curves}  & \colhead{The remaining light curves}  \\ \hline
     \colhead{} & \colhead{} & \colhead{} & \colhead{By white noise$^{a}$} & \colhead{By photometric quality$^{b}$}  & \colhead{} 
     }
     \startdata
     \multirow{3}{*}{SMC} & $g$    & 1941 & 1041 & 298 & 602  \\
                          & $V$    & 1941 & 1476 & 98 & 367  \\
                          & $I$    & 1272 & 33 & 29 & 1210 \\
                          \hline
     \multirow{3}{*}{LMC} & $g$    & 4266 & 1736 & 684 & 1846 \\
                          & $V$    & 4266 & 2357 & 385 & 1524  \\
                          & $I$    & 2445 & 73 & 63 & 2309  \\
     \enddata
     \tablecomments{$^{a}$The number of light curves that are considered as white noise, as described in Section \ref{sec: white_noise}.
     $^{b}$The number of light curves deviating from the acceptable range in Figure \ref{fig: Photomtric_selection}, which are considered to have poor photometric quality. \\
     }
\end{deluxetable}

\begin{deluxetable}{ccccc}[h]
     \tablecaption{The selected range of $Q$ for each band. \label{tab: Q_range}}
     \tablewidth{0pt}
     \tablehead{\colhead{Galaxy}  & \colhead{Band}& \colhead{$Q_{\rm mean}$}& \colhead{$Q_{\rm min}$} & \colhead{$Q_{\rm max}$}}
     \startdata
     \multirow{3}{*}{SMC} & $g$    & 0.783 & 0.209 & 2.930   \\
                          & $V$    & 1.083 & 0.247 & 4.756 \\
                          & $I$    & 0.658 & 0.254 & 1.705\\
                          \hline
     \multirow{3}{*}{LMC} & $g$    & 0.677 & 0.203 & 2.256\\
                          & $V$    & 0.801 & 0.192 & 3.348  \\
                          & $I^{a}$    & 0.658 & 0.254 & 1.705  \\
     \enddata
     \tablecomments{$^{a}$ The same range as the $I$ band of the SMC is selected here.}
 \end{deluxetable}

\begin{sidewaystable}  
     \centering
     \caption{Granulation and stellar parameters of RSGs in the SMC}
     \label{tab:params_smc}
     \begin{tabular}{cccccccccccccccccc}
          \toprule
          R.A. & Decl. & $\mathrm{ID}_g$ & $Q_g$ & $\sigma_g$ & $\rho_g$ & $\tau_g$ & $\sigma_g^{\mathrm{PSD}}$ & $\tau_g^{\mathrm{PSD}}$ & $\alpha_g^{\mathrm{PSD}}$ & $\tau_{\mathrm{eff},g}^{\mathrm{PSD}}$ & $V^{a}$ & $I^{b}$ & $T_{\mathrm{eff}}$ & $\log(L/L_{\odot})$ & $M/M_{\odot}$ & $R/R_{\odot}$ & $\log\ g$ \\
          \midrule
          & & & & (mmag) & (day) & (day) & (mmag) & (day) & & (day) & & & (K) & & & & \\
          \midrule
          5.825166 & -74.390671 & 125 & 0.363 & $21.6_{-1.9}^{+2.4}$ & $50.5_{-11.5}^{+10.1}$ & $5.8_{-2.7}^{+2.9}$ & 15.4 & 11.0 & 2.97 & 16.7 & ... & ... & 4342 & 4.24 & 11.48 & 231 & 0.767 \\
          8.494317 & -72.847710 & 431 & 0.362 & $37.1_{-4.5}^{+6.3}$ & $116.0_{-20.5}^{+23.2}$ & $13.4_{-5.1}^{+6.6}$ & 20.8 & 45.1 & 2.11 & 48.9 & ... & ... & 4089 & 4.57 &13.88 & 382 & 0.415 \\
          8.763859 & -74.203384 & 476 & 0.644 & $28.7_{-2.8}^{+3.5}$ & $88.5_{-10.2}^{+11.7}$ & $18.1_{-5.4}^{+6.9}$ & 15.6 & 10.1 & 4.56 & 18.4 & ... & ... & 4322 & 4.44 & 12.90 & 295 & 0.607 \\
          9.064605 & -73.663338 & 526 & 0.565 & $42.7_{-4.5}^{+5.8}$ & $91.7_{-20.9}^{+16.9}$ & $16.5_{-8.5}^{+10.1}$ & 25.9 & 17.7 & 2.88 & 26.2& ... & ... & 3999 & 4.38 & 12.44 & 321 & 0.519 \\
          \bottomrule
     \end{tabular}
     \tablecomments{$^{a,b}$ $V$ and $I$ bands are provided with the same columns as $g$ band, i.e., $\mathrm{ID}$, $Q$, $\sigma$, $\rho$, $\tau$, etc.\\
     (This table is available in its entirety in machine-readable form.)}
\end{sidewaystable}

\begin{sidewaystable}  
     \centering
     \caption{Granulation and stellar parameters of RSGs in the LMC}
     \label{tab:params_lmc}
     \begin{tabular}{cccccccccccccccccc}
       \toprule
       R.A. & Decl. & $\mathrm{ID}_g$ & $Q_g$ & $\sigma_g$ & $\rho_g$ & $\tau_g$ & $\sigma_g^{\mathrm{PSD}}$ & $\tau_g^{\mathrm{PSD}}$ & $\alpha_g^{\mathrm{PSD}}$ & $\tau_{\mathrm{eff},g}^{\mathrm{PSD}}$ & $V^{a}$ & $I^{b}$ & $T_{\mathrm{eff}}$ & $\log(L/L_{\odot})$ & $M/M_{\odot}$ & $R/R_{\odot}$ & $\log\ g$ \\
       \midrule
       & & & & (mmag) & (day) & (day) & (mmag) & (day) & & (day) & & & (K) & & & & \\
       \midrule
       66.437113 & -69.722 & 93 & 0.726 & $31.1_{-2.7}^{+3.3}$ & $84.8_{-7.4}^{+8.4}$ & $19.6_{-4.5}^{+5.5}$ & 18.5 & 9.1 & 4.23 & 16.3 & ... & ... & 3926 & 4.56 & 13.78 & 408 & 0.354 \\
       68.085530 & -65.023895 & 178 & 0.896 & $23.8_{-2.0}^{+2.1}$ & $23.0_{-2.9}^{+2.4}$ & $6.6_{-2.6}^{+3.0}$ & 11.5 & 3.1 & 11.80 & 6.7 & ... & ... & 4014 & 4.09 & 10.55 & 229 & 0.740 \\
       70.448950 & -60.047489 & 475.0 & 1.105 & $22.4_{-1.5}^{+1.7}$ & $32.1_{-2.5}^{+2.5}$ & $11.3_{-2.9}^{+3.5}$ & 13.2 & 3.8 & 9.39 & 7.8 & ... & ... & 4054 & 4.24 & 11.45 & 264 &  0.651 \\
       70.465616 & -69.837532 & 479.0 & 0.378 & $119.4_{-13.8}^{+18.6}$ & $123.4_{-16.9}^{+21.0}$ & $14.9_{-4.2}^{+5.3}$ & 41.3 & 19.7 & 5.53 & 37.9 & ... & ... & 4069 & 3.88 & 9.35 & 175 &  0.921 \\
       \bottomrule
     \end{tabular}
     \tablecomments{$^{a,b}$ $V$ and $I$ bands are provided with the same columns as $g$ band, i.e., $\mathrm{ID}$, $Q$, $\sigma$, $\rho$, $\tau$, etc.\\
     (This table is available in its entirety in machine-readable form.)}

\end{sidewaystable}

\begin{deluxetable}{ccc}[h] \label{tab: rho_scaling_relations}
	\tablecaption{The scaling relations of $\rho$ with stellar parameters}
	\tablewidth{0pt}
	\tablehead{\colhead{Stellar}  & \multicolumn2c{$\log\ \rho\ \ (\mathrm{day})$}  \\ \hline
     \colhead{paremeters}  & \colhead{SMC} & \colhead{LMC}}
	\startdata
	$\log\ g$  & $-0.971X+2.437^{a}$ & $-1.052X+2.424$ \\
	$M/M_{\odot}$ & $0.100X+0.598$ & $0.106X+0.543$ \\
     $T_{\mathrm{eff}}\ \ (K)$ &  $-0.0015X+8.324$ & $-0.0019X+9.758$ \\
     $\log\ L/L_{\odot}$ & $0.807X-1.675$ & $0.893X-2.036$ \\
     $\log \ R/R_{\odot}$ &  $1.498X-1.841$ & $1.630X-2.210$ \\
	\enddata
     \tablecomments{$^{a}$ $X$ in the scaling relations represents the stellar parameters in the first column, for example, which should be understood here as $\log\ \rho=-0.971\ \log\ g+2.437$.}
\end{deluxetable}

\begin{table}
     \caption{The scaling relations of $\sigma$ with stellar parameters}
     \label{tab: sigma_scaling_relations}
     \centering
     \begin{tabular}{ccc}
         \toprule
         Stellar parameters & \multicolumn{2}{c}{$\log\ \sigma\ \ (\mathrm{mmag})$} \\
         \cline{2-3}
          & SMC & LMC \\
         \midrule
         \multirow{2}{*}{$\log\ g$} & $-1.456X+2.036^{a}$ for $X<0.390$ & $-1.368X+2.052$ for $X < 0.433$ \\
                                    & $-0.512X+1.668$ for $X>0.390$ & $-0.347X+1.610$ for $X > 0.433$ \\
          \midrule
         \multirow{2}{*}{$M/M_{\odot}$} & $0.054X+0.674$ for $X<13.65$ & $0.047X+0.849$ for $X < 12.86$ \\
                                        & $0.121X-0.242$ for $X>13.65$ & $0.119X-0.080$ for $X > 12.86$ \\
          \midrule
          \multirow{2}{*}{$T_{\mathrm{eff}}\ (K)$} & $-0.0021X+10.185$ for $X<4102$ & $-0.0026X+12.006$ for $X<4010$ \\
                                                 & $-0.0012X+6.607$ for $X > 4102$ & $-0.0005X+3.428$ for $X > 4010$ \\
          \midrule
         \multirow{2}{*}{$\log\ L/L_{\odot}$} & $0.325X-0.082$ for $X<4.55$ & $0.285X+0.179$ for $X<4.48$ \\
                                              & $1.191X-4.023$ for $X >4.55$ & $1.158X-3.729$ for $X > 4.48$ \\
          \midrule
          \multirow{2}{*}{$\log \ R/R_{\odot}$} & $0.773X-0.546$ for $X<2.59$ & $0.531X+0.096$ for $X<2.56$ \\
                                               & $2.249X-4.383$ for $X > 2.59$ & $2.118X-3.969$ for $X > 2.56$ \\
         \bottomrule
     \end{tabular}
     \tablecomments{$^{a}$The table is represented in the same way as mentioned in the caption of Table \ref{tab: rho_scaling_relations}.}
 \end{table}


\end{CJK*}
\end{document}